\documentclass[12pt]{iopart}
\usepackage{iopams}  
\usepackage[dvips,dvipdfm]{graphicx}
\newcommand{\Part}[3]{ \frac{ \partial^{#3} #1 }{ \partial #2^{#3} }
}%partial derivative 
\newcommand{\V}[1]{{\mbox{\boldmath $#1$}}} %vector command
 
\newcommand{\ket}[1]{|{#1}\rangle } %ket vector
\newcommand{\braket}[2]{\langle {#1}|{#2}\rangle }
\newcommand{\Ave}[1]{\left\langle {#1} \right\rangle} %thermal average 
\newcommand{\Wt}[1]{ {\widetilde {#1}} } %attachment of  tilde 
\newcommand{\Wh}[1]{\widehat {#1}} %attachment of hat 
\newcommand{\sgn}[1]{{\rm sgn}({#1})} 
\newcommand{\llsim}{\ $\raisebox{-.7ex}{$\stackrel{\textstyle <}{\sim}$}$\,\ }
\newcommand{\gsim}{\ $\raisebox{-.7ex}{$\stackrel{\textstyle >}{\sim}$}$\,\ }

\newcommand{\mC}{\mathbb{C}}
\newcommand{\mR}{\mathbb{R}}
\newcommand{\mN}{\mathbb{N}}

\begin{document}

\title%[Author guidelines for IOPP journals]
{Complex Replica Zeros of $\pm J$ Ising Spin Glass at Zero Temperature}

\author{Tomoyuki Obuchi\dag\footnote[3]{obuchi@stat.phys.titech.ac.jp}, 
Yoshiyuki Kabashima\ddag and Hidetoshi Nishimori\dag
}

\address{
\dag
Department of Physics, Tokyo Institute of Technology,\\  
Tokyo 152-8551, Japan \\
\ddag
Department of Computational Intelligence and Systems Science, \\
Tokyo Institute of Technology, Yokohama 226-8502, Japan\\
}

\begin{abstract}
Zeros of the $n$th moment of the partition function $[Z^n]$ 
are investigated in a vanishing temperature limit $\beta \to \infty$, 
$n \to 0$ keeping $y=\beta n \sim O(1)$. 
In this limit, the moment parameterized by $y$
characterizes the distribution of the ground-state energy. 
We numerically investigate the zeros for $\pm J$ Ising spin glass models 
with tree and other several systems, which 
can be carried out with a feasible computational cost 
by a symbolic operation based on the Bethe--Peierls method. 
For several tree systems we find that the zeros tend to approach 
the real axis of $y$ in the thermodynamic limit
implying that the moment cannot be described by a single 
analytic function of $y$ as the system size tends to infinity, 
which may be associated with breaking of the replica symmetry. 
However, examination of the analytical properties of the 
moment function and assessment of 
the spin-glass susceptibility 
indicate that the breaking of analyticity is relevant to
neither one-step or full replica symmetry breaking. 
\end{abstract}



\maketitle
\section{Introduction}
Spin glasses are a typical example of disordered 
systems and have been investigated
for a long time \cite{SPIN}. The first comprehensive understanding of 
spin glasses 
was obtained by investigating the so-called SK model introduced by Sherrington 
and Kirkpatrick \cite{Sher}, which describes fully
connected Ising spin glasses. 
In analyzing this model, they employed the replica method 
under the replica symmetric (RS) ansatz. 
However, the SK solution contains an inconsistency 
in that the entropy at low temperatures becomes negative. 
This problem has led to much controversy regarding the
validity of the replica method.
In 1980, Parisi developed
the replica symmetry breaking (RSB) scheme \cite{Pari1,Pari2} and showed that 
a sufficient solution 
can be obtained within the framework of the replica method.  
%The inconsistency of the RS solution was overcome by this new scheme and discussions
%about the mathematical treatment of the replica method have abated.

Although Parisi's RSB scheme is consistent 
at low temperatures, a mathematical justification of the replica method 
and a proof of the Parisi scheme were lacking
until a recent study showed that the Parisi's solution is 
exact for the SK model \cite{Tala}.
However, this does not resolve all of the questions regarding the replica method. 
There are still many unsolved issues,  
e.g. ultrametricity and 
the origin of the RSB.  
These issues have attracted renewed interest as applications of the
the replica method have increased rapidly \cite{STAT,Sour,Kaba,Nish}, and
a deeper understanding of this method is greatly desired.
 
The RSB is considered to relate to the analyticity of a
generating function $g(n)$ defined as follows:
\begin{eqnarray}
g(n)\equiv \lim_{N \rightarrow \infty} g_{N}(n), \label{g_n}\\
g_{N}(n)\equiv \frac{1}{N}\log [Z^n], 
\label{g_n2}
\end{eqnarray}
where $n$ is referred to as the replica number and the 
brackets $[\cdots]$
denote the average over the quenched randomness. 
The functions $g_{N}(n)$ and $g(n)$ are defined 
for $\forall{n} \in \mR$ (or $\in \mC$) and the free energy is derived from 
$g_{N}(n)$ as
\begin{equation}
f=-\lim_{N \rightarrow \infty }
\frac{1}{\beta N}[\log Z]=-\lim_{N \rightarrow \infty }
\lim_{n \rightarrow 0}\frac{1}{n\beta } g_{N}(n). \label{eq:the replica method}
\end{equation}
The name `replica method' is often used to indicate 
the second identity, 
though this method should be 
considered as a systematic procedure to evaluate 
eqs.\ (\ref{g_n}) and (\ref{g_n2}). 
In general, the calculation of $[Z^n]$ is 
difficult for real $n \in {\mR }$ (or complex $\mC$). 
To overcome this difficulty, 
the replica method first computes $[Z^n]$ for natural 
numbers $n=1,2,\cdots \in \mN$, then 
extends the obtained expressions of 
$[Z^n]$ to $n \in \mR$ by their analytical continuation. 
However, this technique causes the following two problems. 
The first concerns the uniqueness of 
the analytical continuation from natural to real
numbers. Even if all the moments of $[Z^n]$ are given 
for $n \in \mN$, in general it is impossible to uniquely 
continue the analytical expressions for $n\in \mN$ to 
$n \in {\mR}$ (or $\mC$). Carlson's theorem guarantees that
the analytical continuation from $n \in \mN$
to $n \in \mC$ is uniquely determined 
if $[Z^n]^{1/N}< O(e^{\pi |n| })$  
holds as ${\rm Re}(n)$ tends to infinity \cite{THEO}. 
Unfortunately, the moments of the SK model grow as
$[Z^n]^{1/N}< O(e^{C |n|^2})$, where $C$ is a constant,
and therefore this sufficient condition is not satisfied. 
van Hemmen and Palmer conjectured that 
the failure of the RS solution of the SK model 
might be related to this issue, 
though further exploration in this direction is technically difficult
\cite{Hemm}. 
The second issue concerns the possible breaking of the analyticity 
of $g(n)$. In general, even if $g_N(n)$ is guaranteed to be analytic
with respect to $n$ for finite $N$, the analyticity 
of $g(n)=\lim_{N \to \infty} g_N(n)$
can be broken. Since it is unfeasible to exactly compute $[Z^n]$ except for
a few solvable models, in most cases, only the asymptotic behavior 
is investigated by using certain techniques such as the saddle-point method
in the limit $N \to \infty$. 
This implies that, in such cases, the expression analytically continued from 
$n \in \mN$ to $n \in \mR$ in the limit $N \to \infty$ 
will lead to an incorrect solution for $n \to 0$ if the 
breaking of analyticity occurs in the region $0 < n < 1$. 
Recently, it has been shown 
that analyticity breaking 
does occur and is relevant to one-step RSB
(1RSB) for a variation of %discretely 
discrete random energy models  
\cite{Ogur1,Mouk1,Mouk2,Derr}, 
for which the uniqueness of the analytical continuation is guaranteed
by Carlson's theorem and for which $[Z^n]$ or equivalently 
$g_N(n)$ can be assessed in a feasible 
manner without using the replica method for finite ${N}$ 
and ${n} \in {\mC}$. 
This is a strong motivation to investigate the analyticity of 
$[Z^n]$ for various systems to explore possible links to 
different types of RSB.

Under this motivation, we investigate 
possible scenarios of analyticity breaking of 
$g(n)=\lim_{N \to \infty}g_N(n)$. 
For this purpose, we observe the zeros of $[Z^n]$, 
which will be referred to as ``replica zeros'' (RZs), on the complex plane 
$n \in \mC$ for finite $N$ and examine how 
some sequences of zeros approach the real axis 
as $N$ tends to infinity.

For the discrete random energy model mentioned above, 
this strategy successfully characterizes an RSB 
accompanied by a singularity of a large deviation rate 
function with respect to $N^{-1} \log Z$ \cite{Ogur2}. 
As other tractable example systems, 
we investigate $\pm J$ models 
with a symmetric distribution on two types of lattices, 
ladder systems and Cayley trees (CTs) 
with random fields on the boundary. 
There are two reasons for using these models: 
Firstly, these models can be investigated 
in a feasible computational time by the Bethe--Peierls 
(BP) approach \cite{Bowm}. 
Especially, at zero temperature this approach gives 
a simple iterative formula to yield the
partition function. Employing the replica method and the BP formula, we
can perform symbolic calculations of  
the replicated partition function $[Z^n]$, 
which enables us 
to directly solve the
equation of the RZs $[Z^n]=0$. The second reason is the
existence of the spin-glass phase. It is known that the spin-glass phase is present
for CTs \cite{Chay,Mott,Carl,Lai} 
and is absent for ladder systems. Therefore, we can 
compare the behavior of RZs, which are considered to be 
dependent on the spin-glass ordering.  

This paper consists of five sections. In the next section, we give an 
explanation of our formalism. 
Simple recursive equations to calculate $[Z^n]$ are derived 
in a zero-temperature limit by combining the BP approach and the replica method. 
The relationships of CTs to Bethe lattices (BLs) and 
regular random graphs (RRGs) are also argued. 
Assessing the contribution from the boundary 
indicates that 1RSB does not occur in CTs and BLs while 
it does for RRGs in the thermodynamic limit when the boundary 
contribution is correctly taken into account, 
which is the case in the evaluation of RZs. 
This implies that the possible RZs of a CT are irrelevant to 1RSB. 
In sec. 3, we present plots of RZs and investigate their 
behavior. Their physical significance is also discussed.
In sec. 4, a possible link to another type of RSB, the full RSB (FRSB), is 
examined. Numerical assessment of the de Almeida--Thouless (AT)
condition based on the divergence of spin-glass susceptibility, however, 
indicates that RZs do not reflect FRSB, either. 
Therefore, we conclude that the analyticity breaking 
that occurs in CTs is irrelevant to RSB. 
The final section is devoted to a summary. 
   
\section{Formulation}\label{sec:formulation}
In this section, the main ideas of the paper are presented. 
It is shown that the RZ equation $[Z^n]=0$ is simplified at
zero temperature. 
An algorithm to evaluate the generalized moment $[Z^n]$ for 
$n \in \mC$ is developed by introducing the replica method to the BP approach. 
\subsection{Equation of the replica zeros at zero temperature}
\label{sec:aboutT0}
Solving 
\begin{equation}
[Z^n]=0 \label{eq:RZs}
\end{equation}
with respect to $n$ is our main objective. 
Unfortunately, this is, in general, a hard task even by 
numerical methods because eq.\ (\ref{eq:RZs}) is 
transcendental and becomes highly complicated as
the system size $N$ grows.
In the $T\rightarrow 0$ limit, however, the main contributions 
to the partition function come
from the ground state and eq.\ (\ref{eq:RZs}) becomes
\begin{equation}
[Z^n]\approx[d_{ {\rm g} }^{n}e^{-\beta n E_{ {\rm g} }} ]=0,
\label{eq:RZs2}
\end{equation}
where $E_{{\rm g}}$ is the
energy of the ground state and $d_{{\rm g}}$ is the degeneracy. 
If $n$ is finite when $\beta \rightarrow \infty$, the term 
$e^{-\beta n E_{ {\rm g} }}$ diverges or vanishes and there is
no meaningful result. Therefore, we suppose that non-trivial
solutions exist only in the limit 
$n \rightarrow 0,\beta \rightarrow \infty,$ and 
$ y=\beta n \sim O(1)$. 
This assumption is consistent with the 
fact that the solution of the SK model is well-defined 
in this limit \cite{Pari3}.
Under this condition, eq.\ (\ref{eq:RZs2}) becomes 
\begin{equation}
[e^{-yE_{ {\rm g} }} ]=0.
\label{eq:RZs3}
\end{equation}
In the following, 
we focus on the $\pm J$ model whose Hamiltonian is given by
\begin{equation}
H=-\sum_{\Ave{i,j}}J_{ij}S_{i}S_{j},
\end{equation}
and the distribution of interactions is  
\begin{equation}
%P(J_{ij})=p_J\delta(J_{ij}-1)+(1-p_J)\delta(J_{ij}+1), 
P(J_{ij})=\frac{1}{2}\delta(J_{ij}-1)+\frac{1}{2}\delta(J_{ij}+1), 
\label{eq:dist_int}
\end{equation}
assuming that the total number, $N_{B}$, of interacting spin pairs
$\Ave{i,j}$ is proportional to $N$, which is 
the case for ladder systems and CTs. 
This limitation restricts the energy of any state to an integer value.
As a result, eq.\ (\ref{eq:RZs3}) can always
be expressed as a polynomial of $x=e^y$, 
which significantly reduces the numerical cost for 
searching for RZs. 

%%%%%%%%%%%%%%%%%%%%%%%%%%%%%%%%%%%%%%%%
One issue may be noteworthy here. In the present study, we focus on 
the limit $n \to 0$, $\beta \to \infty$ keeping $\beta n \to y \sim O(1)$. 
In research on zeros of partition functions, 
on the other hand, 
another limit $n \to 0$ keeping $\beta$ finite 
can be examined as well. 
In the latter case, the zeros with respect to 
complex $\beta$ are sometimes referred 
to as ``Fisher zeros'' \cite{Fish}. 
Intuitively, Fisher zeros characterize the origin of 
singularities with respect to $\beta$ for typical {\em single sample}
systems. These can be examined not only for 
random systems \cite{Mouk1,Mouk2} but also for systems of 
deterministic interactions such as frustrated anti-ferromagnetic 
Ising spin models \cite{Matv}. As $n \to 0$ limit is taken 
on ahead for each $\beta$, 
Fisher zeros are irrelevant to the analyticity concerning 
the replica number $n$. 
For examination of the analyticity with respect to $n$, 
it is necessary to investigate the zeros of $\left [Z^n \right ]$ 
in the complex plane of $n$. 
In the situation of vanishing temperatures $\beta \to \infty$, 
this naturally leads to the current nontrivial limit
$n \to 0$, $\beta \to \infty$ keeping $\beta n \to y \sim O(1)$. 
%%%%%%%%%%%%%%%%%%%%%%%%%%%%%%%%%%%%%%%%

\subsection{The Bethe--Peierls approach}
\subsubsection{General formula}
On cycle free graphs, it is possible to assess the partition 
function by an iterative method, {\em i.e.} the BP approach. 
We here present a brief review of the procedure 
for CTs. The BP approach in ladder systems is presented in \ref{sec:BPladder}.

\begin{figure}[htbp]
%\hspace{-5mm}
%\begin{minipage}{0.5\hsize}
\begin{center}
   \includegraphics[height=40mm,width=50mm]{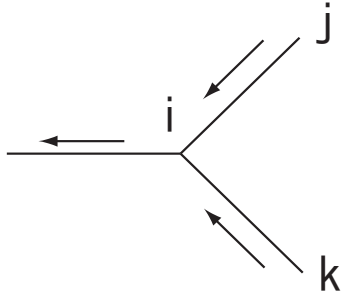}
 \caption{Local structure of a CT with coordination
 number $c=3$.}
 \label{fig:RBG}
\end{center}
\end{figure}

The basis for our analysis is a formula for evaluating an effective field by a partial trace: 
\begin{equation}
\sum_{S_{j}} \exp\left\{
\beta( J_{ij}S_{i}S_{j}+h_{j}S_{j})
\right\}=A\exp(\beta h_{i}S_{i}).
\label{eq:basic}
\end{equation}
A simple algebra offers
\begin{equation}
h_{i}={\widehat h}_{j},\,\, A=\frac{2\cosh \beta J_{ij}\cosh \beta h_{j}}
{\cosh\beta {\widehat h}_{j}}, \label{eq:A}
\end{equation}
where
\begin{equation}
\beta {\widehat h}_{j}=\tanh^{-1}(\tanh \beta J_{ij} \tanh \beta h_{j} ) \label{eq:bias}. 
\end{equation}
The fields $h_{j}$ and $\Wh{h}_{j}$ are sometimes termed the cavity field and 
cavity bias, respectively. For CTs, 
iterating the above equations from the boundary gives 
the series of cavity fields and biases  
$\{ h_{j},\Wh{h}_{j}\}$.
% This procedure leads to
%generations in the tree. Spins traced
%out ahead of the site $i$ are called $i$'s ascendants and spins after the site
%$i$ are $i$'s descendants (ee fig.\ \ref{fig:tree}).  
%\begin{figure}[htbp]
%\begin{center}
%   \includegraphics[height=50mm,width=80mm]{bethelattice.eps}
% \caption{Local structure of a BL with coordination
% number $c=3$.}
% \label{fig:tree}
%\end{center}
%\end{figure}
In general, a cavity field becomes a summation of the
cavity biases from 
its $c-1$ descendants ($c$ is the coordination number): 
\begin{equation}
h_{i}=\sum_{j=1}^{c-1} {\widehat h}_{j}.
\label{eq:bias_to_field}
\end{equation}
Hereafter, we mainly focus on the $c=3$ case, as shown in fig.\ \ref{fig:RBG}, 
but the extension to general coordination numbers is straightforward.
%%%
%%%
%%%
In addition, generalizing to $k$-spin interacting CTs ($k$-CTs)
is also straightforward;
the only necessity is to replace the partial trace (\ref{eq:basic}) 
with that for a $k$-spin interaction, as 
\begin{equation}
\sum_{S_{1},S_{2}} \exp
\left\{
\beta\left( S_{i}J_{k}\prod_{j=1}^{k-1}S_{j}+\sum_{j=1}^{k} h_{j}S_{j}
\right)
\right\}
=A\exp(\beta h_{i}S_{i}),
\label{eq:3bbasic}
\end{equation}
where
\begin{equation}
h_{i}=\Wh{h}_{k},\,\, A=\frac{2^{k-1}\cosh\beta J_{k} \prod_{j=1}^{k-1}\cosh\beta h_{j}}{\cosh \beta \Wh{h}_{k}},
\end{equation}
\begin{equation}
\Wh{h}_{k}=\frac{1}{\beta}\tanh^{-1}\left(
\tanh \beta J_{k} \prod_{j=1}^{k-1}\tanh \beta h_{j}
\right).
\end{equation}
%The $3$-CT's structure with $c=3$ is also shown in fig.\ \ref{fig:3bRBG}.
%%%%%
%%%%%

Let us denote the partition function in the absence of $i$'s ascendants as $Z_{i}$. 
Equations (\ref{eq:basic})--(\ref{eq:bias_to_field})
imply that the partition function is updated as 
\begin{equation}
Z_{i}=
\sum_{S_{i},S_{j},S_{k}}Z_{j}Z_{k}\exp\{-\beta (\Delta H_{ij}+\Delta H_{ik})\} 
\rho_{j}(S_{j},h_j)
\rho_{k}(S_{k},h_k),\label{eq:BPgeneral}
\end{equation}
where
\begin{equation}
\rho_{j}(S_{j})=\frac{\exp(\beta h_{j}S_{j})}{2\cosh \beta h_{j}},
\end{equation}
is the one-site marginal in the absence of $j$'s ascendants and
$\Delta H_{ij}=-J_{ij}S_{i}S_{j}$
is the bond Hamiltonian added by a propagation procedure.

As a final step, the contribution from the origin of 
the tree is calculated as  
\begin{eqnarray}
&&\hspace{-25mm}Z= 
Z_{1}Z_{2}Z_{3}\prod_{i=1}^{3}(2\cosh\beta J_{i}) \cr %\notag \\
&&\hspace{-25mm} \times \left(
\frac{1+\tanh\beta J_{1}\tanh\beta J_{2}\tanh\beta h_{1}\tanh\beta h_{2}+
({\rm two \,\, terms\,\, with\,\, 1,2,3\,\, rotated}) }{4}
\right), 
\end{eqnarray}
and the whole partition function $Z$ is derived. 
Taking the $T \rightarrow 0$ limit yields the ground-state
energy for a given bond configuration. 
In this limit, eqs.\ (\ref{eq:bias}) and %(\ref{eq:Zi2})
(\ref{eq:BPgeneral})
become
\begin{equation}
{\widehat h}_{j} \rightarrow \sgn{J_{ij}h_{j}}\min(|J_{ij}|,|h_{j}|), \label{eq:T0bias}
\end{equation}
and
\begin{eqnarray}
&&\hspace{-25mm} \lim_{\beta \rightarrow \infty} -\frac{1}{\beta}\log Z_{i}=E_{i} \cr %\notag \\
&&\hspace{-25mm} =E_{j}+E_{k}
-J_{ij}-J_{ik}+
\left\{
\begin{array}{ll}
0 & \,\, (\,\,\sgn{J_{ij}J_{ik}h_{j}h_{k}} \geq 0\,\,) \,\,\\
2\min(|J_{ij}|,|J_{ik}|,|h_{j}|,|h_{k}|) & \,\, (\,\,{\rm otherwise}\,\,)\,\, 
\end{array}
\right..
\end{eqnarray}
We assume $\sgn{0}\equiv 0$ in this paper.

\subsubsection{Use of the replica method}
For a given single sample of interactions and boundary conditions, 
a simple application of the BP algorithm enables us to 
evaluate the partition function in a feasible computational time. 
Unfortunately, this does not fully resolve the problem of the computational 
cost for assessing the moments (\ref{eq:RZs}) since 
the cost for evaluating an average over all possible samples of 
interactions and boundary conditions grows exponentially with 
respect to the number of spins. However, this difficulty can be 
overcome by analytically assessing the configurational average 
for $n \in \mN$  and analytically continuing the obtained 
expressions to $n \in \mC$ in the level of the algorithm, 
a method which may be considered a generalization of the replica method. 

For this purpose, we first evaluate the $n$th moment of the 
partition function $Z$ 
\begin{eqnarray}
&&\Xi(n)\equiv [Z^n]=\Tr{} \prod_{\Ave{i,j} } [\exp(\beta J_{ij}\sum_{\alpha}^{n}
S^{\alpha}_{i}S^{\alpha}_{j}  ) ], \label{eq:effZ}
\end{eqnarray}
for $n \in \mN$, where $\alpha$ is the replica index.  
Let us denote the effective Hamiltonian as
\begin{equation}
H_{{\rm eff}}
=\sum_{\Ave{i,j}} H_{ij}
=\sum_{\Ave{i,j}}-\frac{1}{\beta}\log [ \exp(\beta J_{ij} \sum_{\alpha}S_{i}^{\alpha}S_{j}^{\alpha})],
\end{equation}
where $[\cdots]$ stands for the configurational average with respect to 
the interactions $\{J_{ij}\}$. 
This means that eq.\ (\ref{eq:effZ}) is simply 
the partition function of an $n$-replicated system,
which is defined on a cycle free graph and is free from 
quenched randomness. Therefore, by expressing the BP algorithm in the 
current case as 
\begin{eqnarray}
&&\hspace{-15mm} \Xi_{i}(n)=
\sum_{\V{S}_{i}, \V{S}_{j}, \V{S}_{k}} 
\left[\exp\left\{\beta (J_{ij}\sum_{\alpha}^{n}
S^{\alpha}_{i}S^{\alpha}_{j} +J_{ik}\sum_{\alpha}^{n}
S^{\alpha}_{i}S^{\alpha}_{k})\right\}
\right]
\rho_{j} \rho_{k }  \Xi_{j}(n) \Xi_{k}(n)  \label{eq:SPijk}\\
&&\hspace{-15mm} 
=\sum_{\V{S}_{i}} 
\rho_{i}(\V{S}_{i})  \Xi_{i}(n),\label{eq:SPi} 
\end{eqnarray}
where $\rho_{i}$ is the one-site marginal distribution 
of site $i$, eq.\ (\ref{eq:effZ}) can be assessed in a 
feasible time. The expressions (\ref{eq:SPijk}) and (\ref{eq:SPi})
define the updating rules of $\rho_{i}$ and $\Xi_{i}(n)$. 

So far, we have made no assumptions or approximations
and therefore eq.\ (\ref{eq:SPi}) yields exact assessments 
for $n \in \mN$, given a boundary condition. 
To generalize this scheme to $n \in \mC$, 
we here introduce the RS ansatz, which is the second step of the replica method and, 
in general, is expressed by a restriction of the functional 
form of $\rho_{i}(\V{S}_{i})$ as 
\begin{equation}
\rho_{i}(\V{S}_{i})=
\int \pi_{i}(h)\prod_{\alpha=1}^{n}
\left(\frac{1+\tanh(\beta h) S^{\alpha }_{i} }{2}  
\right)dh
=\int \pi_{i}(h)
\frac{e^{\beta h \sum_{\alpha} S^{\alpha}_{i}  }  }{(2\cosh \beta h)^n}  dh, 
\label{eq:pifi}
\end{equation}
where $\pi_i(h)$ is a distribution to be updated in the algorithm. 
The expression of eq.\ (\ref{eq:pifi}) guarantees that 
$\rho_i(\V{S}_{i})$ is invariant under any permutation of the replica 
indices $\alpha=1,2,\ldots,n$. Note that this property is 
automatically satisfied over all of the objective lattice if only 
the distributions on the boundary 
are expressed in the form of eq.\ (\ref{eq:pifi}). 

Inserting eq.\ (\ref{eq:pifi}) into eq.\ (\ref{eq:SPi})
and performing some simple algebraic steps gives 
\begin{eqnarray}
&& \Xi_{i}=\sum_{ \V{S}_{i}}
\Xi_{j}\Xi_{k}
( 2\cosh\beta)^{2n}\int dh_{i}
\frac{e^{\beta h_{i} \sum_{\alpha} S^{\alpha}_{i}}}{(2\cosh \beta h_{i})^n} 
\Bigg\{ 
\int \!\!\! \int \pi_{j}(h_{j}) \pi_{k}(h_{k}) \cr %\notag  \\
&& \hspace{-0cm}\times \left[
\delta(h_{i}-{\widehat h}_{j}-{\widehat h}_{k} ) 
\left(
\frac{2\cosh \beta h_{i}}{2\cosh\beta {\widehat h}_{j}2\cosh\beta {\widehat h}_{k} }
\right)^n
\right]dh_{j}dh_{k} 
\Bigg\} \label{eq:rhoijZij} \\
&&=
\sum_{ \V{S}_{i}}
\int dh_{i}
\pi_{i}(h_{i}) 
\frac{e^{\beta h_{i} \sum_{\alpha} S^{\alpha}_{i}}}{(2\cosh \beta h_{i})^n} 
\Xi_{i}. 
\label{eq:rhoiZi}
\end{eqnarray}
Equations (\ref{eq:rhoijZij}) and (\ref{eq:rhoiZi}) provide an 
expression of the replica symmetric BP algorithm: 
\begin{eqnarray}
\hspace{-15mm} \pi_{i}(h_{i})
\propto
\int \!\!\! \int \pi_{j}(h_{j}) \pi_{k}(h_{k})\left[
\delta(h_{i}-{\widehat h}_{j}-{\widehat h}_{k} ) 
\left(
\frac{2\cosh \beta h_{i}}{2\cosh\beta {\widehat h}_{j}2\cosh\beta {\widehat h}_{k} }
\right)^n
\right]dh_{j}dh_{k}\label{eq:pih1},  
\end{eqnarray}
\begin{eqnarray}
\hspace{-15mm} \Xi_i=\Xi_{j}\Xi_{k}( 2\cosh\beta)^{2n}
\int \!\!\! \int dh_{j}dh_{k}
\, \pi_{j}(h_{j})\pi_{k}(h_{k}) 
\left[
\left(
\frac{2 \cosh \beta (\widehat{h}_j +\widehat{h}_k)}{
2\cosh\beta {\widehat h}_{j}2\cosh\beta {\widehat h}_{k} }
\right) ^n
\right],\label{eq:Zbranch}
\end{eqnarray}
which is applicable to $\forall{n} \in \mC$. 
When the algorithm reaches the origin of the CT, 
the moment of eq.\ (\ref{eq:effZ}) is assessed as 
\begin{eqnarray}
&&\Xi(n)=[Z^n]
=\Xi_{1}\Xi_{2}\Xi_{3}(2\cosh\beta )^{3n}
\int \!\!\! \int\!\!\! \int dh_{1}dh_{2}dh_{3}
\, \pi_{1}(h_{1})\pi_{2}(h_{2})\pi_{3}(h_{3})  \cr %\notag \\
&& \times\left[
\left(
\frac{1+\tanh\beta J_{1}\tanh\beta J_{2}\tanh\beta h_{1}\tanh\beta h_{2}
+R }{4}
\right) ^n
\right],\label{eq:Zfinal}
\end{eqnarray}
where $R$ is two terms with the indices $1,2,3$ rotated.

\subsubsection{Zero-temperature limit}
Under appropriate boundary conditions, the zero-temperature limit $\beta \to \infty$, $n \to 0$ 
keeping $y=\beta n$ finite, which we focus on in the present paper, 
yields further simplified expressions of the BP algorithm. 
For this, we generate replicated spins of each site on 
the boundary with an identical random external field $h_i=\pm 1$,
the sign of which is determined with an equal probability of $1/2$. 
This yields the cavity field distribution
\begin{eqnarray}
\pi_i(h_i)=\frac{1}{2} \left ( \delta(h_i-1)+\delta(h_i+1) \right )
\label{eq:boundary_dist}
\end{eqnarray}
and the partition function 
\begin{eqnarray}
\Xi_{i}=(2 \cosh \beta)^n \to e^y, 
\label{eq:boundary_Z}
\end{eqnarray}
as the boundary condition. 
The relevance of the boundary condition to the current objective systems 
is discussed later. 

Equation (\ref{eq:boundary_dist}) in conjunction with 
the property $|J_{ij}|=1$ allows
$\pi_i(h_i)$ in eq.\ (\ref{eq:pih1}) to be expressed without loss of generality as 
\begin{equation}
\pi_{i}(h_{i})=
p_{i;0}\delta(h_i)+
\sum_{f=1}^{c-1}
p_{i;f}\left (\delta(h_{i}-f)+\delta(h_{i}+f) \right ),
\label{eq:symmdist}
\end{equation}
where
$\V{p}_i=(p_{i;0}, p_{i;1}, \ldots, p_{i;c-1})$ represents 
a probability vector satisfying $p_{i;0}+2 \sum_{f=1}^{c-1} p_{i;f}=1$
and $p_{i;f}\ge 0$ $(f=0,1,\ldots,c-1)$, 
and is to be determined from the descendent distributions. 
It is noteworthy that the symmetry $\rho_i(\V{S}_i)=\rho_i(-\V{S}_i)$
on the boundary condition also restricts 
$\pi_i(h_i)$ to a symmetric function of the form of 
eq.\ (\ref{eq:symmdist}). 

After the configurational average is performed, the cavity-field 
distribution $\pi_{i}(h_{i})$ depends only on the 
distance, $g$, from the boundary. Therefore, we hereafter denote 
$\pi_{i}(h_{i})$ as $\pi_{g}(h_{i})$ and represent 
the distance of the origin from the boundary as $g=L$. 
The BP scheme assesses $\V{p}_{g+1}$
using its descendents $\V{p}_{g}$. 
However, the only part relevant to the assessment of $\Xi(n)$ is 
that for $p_{g;0}$, which is represented as
\begin{equation}
p_{g+1;0}=\frac{p_{g;0}^2+2\left( \frac{1-p_{g;0}}{2} \right)^2 
e^{-2y}}{1-2(1-e^{-2y})\left( \frac{1-p_{g;0}}{2}\right)^2}, 
\label{eq:p0precise}
\end{equation}
for the $c=3$ case, being accompanied by an update of the partition function 
\begin{eqnarray}
\Xi_{g+1}=\Xi_{g}^2e^{2y}
\left\{
1-
2(1-e^{-2y})\left(
\frac{1-p_{g;0}}{2}
\right)^2
\right\}\label{eq:T0Zh1}, 
\end{eqnarray}
and similarly for a general $c$. 
After evaluating $p_{g;0}$ and $\Xi_g$ using this algorithm 
up to $g=L-1$, the full partition function, $\Xi(y)$, in the current limit
$n \to 0$ and $\beta \to \infty$ keeping $y=n\beta \sim O(1)$
is finally assessed as
\begin{eqnarray}
\Xi(y)=\Xi_{L}=\Xi_{L-1}^{3}e^{3y}
\left\{
1-3(1-e^{-2y})\left( \frac{1-p_{L-1;0}}{2} \right)^2(1+p_{L-1;0})
\right\}. 
\label{eq:T0Zh2}
\end{eqnarray}

%%%%%%%%%%%%%%%%%%%%
For $\forall{y} \in \mC$, 
eqs.\ (\ref{eq:p0precise})--(\ref{eq:T0Zh2}) can be performed 
in a feasible computational time
and therefore offer a useful scheme for examining RZs.
This is the main result of the present paper.  
The concrete procedure to obtain RZs is summarized as follows:
\begin{enumerate}
\item{To obtain a series of $p_{g;0}$, eq. (\ref{eq:p0precise}) 
is recursively applied under the initial condition $p_{0;0}=0$ until 
$g$ reaches $L-1$. This can be symbolically performed by using 
computer algebra systems such as {\em Mathematica}.} 
\item{Using the series $\{ p_{g;0} \}$, the moment $\Xi_{g}$ is recursively 
calculated by using eq. (\ref{eq:T0Zh1}) under the initial 
condition (\ref{eq:boundary_Z}) until $g$ becomes $L-1$. 
Then, the full moment $\Xi(y)=\Xi_{L}$ 
is derived from eq. (\ref{eq:T0Zh2}) using $\Xi_{L-1}$ and $p_{L-1;0}$.}
\item{Solving $\Xi_{L}=0$ with respect to $x=e^{y}$ numerically.}
\end{enumerate} 

Although the right hand side of eq. (\ref{eq:p0precise}) 
is expressed as a rational function, $\Xi_{L}$ and $\Xi(y)$ are
guaranteed to be certain polynomials of $x$ since 
the contribution from the denominator is canceled 
in each step of eqs. (\ref{eq:T0Zh1}) and (\ref{eq:T0Zh2}). 
The procedure of (i), (ii) and (iii) can be performed in a polynomial 
time with respect to the number of spins. 
However, for CTs the number of spins and the degree of the polynomial 
$\Xi_{L}$ increase exponentially as $O\left (((k-1)(c-1))^L \right )$ 
as $L$ becomes larger, 
which makes it infeasible to solve $\Xi_{L}=0$ for large $L$. 
For instance, it is computationally difficult
to evaluate RZs beyond $L =7$ for $(k,c)=(2,3)$ and $L=4$ for $(k,c)=(3,4)$
by use of today's computers of reasonable performance. 
This prevents us from accurately examining the convergence of RZs to the real 
axis in the limit $L \to \infty$ by means of numerical methods
and analytical investigation for this purpose is non-trivial either. 
However, the data of small $L$ still strongly indicate that 
the qualitative behavior of RZs
can be classified distinctly 
depending on whether certain bifurcations, which are irrelevant 
to any RSB, occur for the cavity field distribution in the 
limit of $L \to \infty$. 
This implies that RZs of the ladder and tree systems are related to no RSB. 
In the following sections, we give detailed discussions to lead 
this conclusion presenting plots of RZs. 
%%%%%%%%%%%%%%%%%%%

%%%%%%%%%%%%%%%%%%%%%%%%%%%%%%%%%%%%%%%%%%%%%%%%%%%%%%%%%%%%%%%%%%%%%%%%%%%%%%%%%%%%%%%%%%%%%%%%%%%%%%%%%%%%%%%%%%%%%%%%%%%%%%%%%%%%%%%%%%%%%%%%%%%%%%%%%%%%%%%%%%%%%%%%%%%%%%%%%%%%%%%%%%%%%%%%%%%%%%%%%%%%%%%%%%%%%%%%%%%%%%%%%%%%%%%%%%%%%%%%%%%%%%%%%%%%%%%%%%%%%%%%%%%%%%%%%%%%%%%%%%%%%%%%%%%%%%%%%%%%%%%%%%%%%%%%%%%%%%%%%%%%%%%%%%%%%%%%%%%%%%%%%%%%%%%%%%%%%%%%%%%%%%%%%%%%%%%%%%%%%%%%%%%%%%%%%%%%%%%%%%%%%%%%%%%%%%%%%%%%%%%%%%%%%%%%%%%%%%%%%%%%%%%%%%%%%%%%%%%%%%%%%%%%%%%%%%%%%%%%%%%%%%%%%%%%%%%%%%%%%%%%%%%%%%%%%%%%%%%%%%%%%%%%%%%%%%%%%%%%%%%%%%%%%%%%%%%%%%%%%%%%%%%%%%%%%%%%%%%%%%%%%%%%%%%%%%%%%%%%%%%%%%%%%%%%%%%%%%%%%%%%%%%%%%%%%%%%%%%%%%%%%%%%%%%%%%%%%%%%%%%%%%%%%%%%%%%%%%%%%%%%%%%%%%%%%%%%%%%%%%%%%%%%%%%%%%

\subsection{Remarks}
Before proceeding further, there are several issues to be noted. 
\subsubsection{Uniqueness of the analytical continuation}
As already mentioned, analytical continuation 
from $n \in \mN$ to $n \in \mC$ cannot be determined uniquely
in general systems.  
However, in the present system, we can show the uniqueness 
of the continuation. Therefore, the RS solution assumed above
is correct. 

For this, let us consider the modified moment
$[(Z e^{-\beta N_{B}})^n]^{1/N}$, where $N_{B}$ is the total number of
bonds. 
This quantity satisfies the inequality 
\begin{eqnarray}
&&\left|
\left[
\left(
Z e^{-\beta N_{B}}
\right)^n
\right]^{1/N}
\right|
\leq
\left[
\left(
Z e^{-\beta N_{B}}
\right)^{{\rm Re}(n)}
\right]^{1/N} \cr
%&& =
%\left[
%\left(
%\Tr{} 
%\left(
%\prod_{\Ave{i,j}}e^{\beta J_{ij}S_{i}S_{j}} 
%\right)
%e^{-\beta N_{B}}
%\right)^{{\rm Re}(n)}
%\right]^{1/N}
%\cr %\notag \\
&& \leq 
\left[
\left(\Tr{}1
\right)^{{\rm Re}(n)}
\right]^{1/N}
=2^{{\rm Re}(n)}< O(e^{\pi |n| }),
\end{eqnarray}   
for finite $N$.
Suppose that we have an analytic
function $\psi(n;N)$ that satisfies the condition 
$|\psi (n;N)|< O(e^{\pi |n|})$. 
Carlson's theorem guarantees that if the equality 
$|\psi (n;N)-[(Z e^{-\beta N_{B}})^n]^{1/N}|=0$ holds 
for $\forall{n}\in \mN$,
$\psi(n;N)$ is identical to $[(Z e^{-\beta N_{B}})^n]^{1/N}$
for $\forall{n} \in \mC$.   
Because $e^{-\beta N_{B}}$ is a non-vanishing constant, this means
that the analytic continuation of $[Z^{n}]^{1/N}$ is uniquely
determined. This indicates that expressions analytically continued 
under the RS ansatz, namely eqs.\ (\ref{eq:Zfinal}) and (\ref{eq:T0Zh2}), 
are correct for finite $N$ (or equivalently, finite $L$) 
although the analyticity may be broken on the real axis 
in the limit $N \to \infty$. 

\subsubsection{Relationship to other systems} 
In addition to examining RZs for finite CTs and ladder systems, 
the relevance of RZs to the large system size limit will also be 
argued by comparison with known thermodynamic properties of 
relatives of CTs, namely Bethe lattices and regular random graphs. 
These are sometimes identified with CTs because the fixed point 
condition of the BP method is represented identically. 
However, we here strictly distinguish them. 
The definitions and properties of these systems are summarized as follows:
\begin{itemize}
\item The Cayley tree (CT): A tree of finite size consisting of an origin 
      and its neighbors. The first generation is built from $c$ neighbors which
      are connected to the origin. Each site in the $n$th generation is
      connected to new $c-1$ sites without overlap 
      and all these new sites comprise the
      $n+1$th generation. Iterating this procedure to the $L$th
      generation, we obtain the CT, and the $L$th generation 
      becomes its boundary. 
      For the boundary condition of eq.\ (\ref{eq:boundary_dist}), 
      which implies $p_{0;0}=0$ in the expression of 
      eq.\ (\ref{eq:p0precise}), $\Xi(n)$ of this lattice 
      is represented as a polynomial of $x=e^{-y}$, which 
      can be assessed by symbolic operations 
      using eqs.\ (\ref{eq:p0precise})--(\ref{eq:T0Zh2}) without 
      evaluating the values of $\Xi(n)$.  
      This property is very useful for investigating RZs. 

\item The Bethe lattice (BL): A lattice consisting of the first 
  $L^\prime$ generations
      of a CT, for which $L \to \infty$ is taken. 
      Alternatively, we can define a BL as
      a finite CT of $L^\prime$ generation, the boundary condition 
      of which is given by the convergent cavity field distribution 
      of the infinite CT. Unlike for a CT, the boundary condition depends on 
      $y$ for a BL. 
      Due to this difference, $\Xi(n)$ of this lattice cannot be
      represented as a polynomial and searching RZs becomes
      non-trivial. However, assessing the values of 
      $\Xi(n)$ is still feasible computationally.  

\item The regular random graph (RRG): A randomly generated graph 
      under the constraint of a fixed connectivity $c$.
      Since there exist many cycles, assessing $\Xi(n)$ and RZs for this lattice
      is not feasible computationally for finite $N$. 
      In the limit $N \to \infty$ under appropriate conditions, 
      however, it is considered that the RRG and the BL share many
      identical properties. 
      Therefore, this lattice is sometimes identified with 
      the BL and regarded as a solvable system
      \cite{Wong,Meza2,Meza3,Mont}. 
      Nevertheless, we here distinguish between the two systems
      because 
      the main purpose of this paper is to clarify the
      asymptotic properties of 
      $g_{N}(n)$
      from finite $N$ to infinite $N$, and our definition of 
      the BL is useful to compare these limits.
      Here, the terminology ``RRG'' is used only 
      to refer to systems of infinite size. 
\end{itemize} 

\subsubsection{Relevance of the boundary condition to the moment 
of the partition function} 
The above mentioned distinction between the three relatives of CTs
yields differences in the expression of the moment 
of the partition function, even while they share an identical 
cavity field distribution in the limit $N \to \infty$. 

Equations (\ref{eq:p0precise})--(\ref{eq:T0Zh2}) imply that 
$g_N(y)=N^{-1} \log \Xi(y) $ for CTs
is generally expressed as
\begin{eqnarray}
g_N(y)=\frac{1}{N} 
\sum_{\left \langle ij \right \rangle}
g_{\left \langle ij \right \rangle}^{(2)}(y) 
-\frac{1}{N}\sum_{i}(c_i-1) g_{i}^{(1)}(y)
+\frac{1}{N} 
\sum_{\mu}g_{\mu}(y)
, 
\label{eq:decomposition}
\end{eqnarray}
where $g_{\left \langle ij \right \rangle}^{(2)}(y)$
and $g_{i}^{(1)}(y)$ denote the contributions 
from the bond $\left \langle ij \right \rangle$ and 
the site $i$, respectively, and 
$c_i$ is the number of bonds that site $i$ has.
The last term $g_{\mu}(y)$ is the contribution 
due to the boundary fields.
 
This is considered a generalization of a well-known 
property of free energies for cycle free graphs 
\cite{Katsura1979,Nakanishi1981,Yedi}. 
For regular CTs, $c_i=c$ holds if $i$ is placed 
inside the tree, while $c_i=1$ for the boundary sites. 

For a BL, the boundary condition given by 
the convergent solution of eq.\ (\ref{eq:p0precise}), 
$p_*$, which becomes a function of $y$, 
particularly simplifies the expression of 
eq.\ (\ref{eq:decomposition}) as
\begin{eqnarray}
g_N^{\rm BL}(y)=r_{\rm I} g_{\rm I}(y)
+r_{\rm B}g_{\rm B}(y). 
\label{eq:BLdecomposition}
\end{eqnarray}
Here, 
$r_{\rm I}=\left (1+c(c-2)^{-1}
\left ((c-1)^{L^\prime-1}-1 \right ) \right )/
\left (1+c(c-2)^{-1}
\left ((c-1)^{L^\prime}-1 \right ) \right )$ and 
$r_{\rm B}=1-r_{\rm I}$ represent 
the fractions of the number of sites inside the tree and 
on the boundary, respectively, and 
\begin{eqnarray}
g_{\rm I}(y)=\frac{c}{2}g^{(2)}(y)-(c-1)g^{(1)}(y), 
\label{eq:g2g1}
\end{eqnarray}
and
\begin{eqnarray}
g_{\rm B}(y)=\frac{c}{2}g^{(2)}(y)+g_{\mu}(y), 
\label{eq:g2}
\end{eqnarray}
represent contributions from a single site 
inside the tree and on the boundary. In general, 
$g^{(2)}(y)$ and $g^{(1)}(y)$ are expressed as 
\begin{eqnarray}
&&g^{(2)}(y)=\log \left\{ \Tr{}
\left[\Wh{\rho}(\V{S}_{1})^{c-1}\Wh{\rho}(\V{S}_{2})^{c-1}
e^{\beta J\sum_{\alpha}S_{1}^{\alpha}S_{2}^{\alpha}} 
\right]
\right\},\\
&&g^{(1)}(y)=\log \left\{ \Tr{}\Wh{\rho}(\V{S})^{c} \right\},
\end{eqnarray}
where
\begin{equation}
\Wh{\rho}( \V{S} )=\int d \Wh {h } \Wh{ \pi }( \Wh{h} )
\frac{ e^{\beta \Wh{h} \sum_{\alpha} S_{\alpha}  }   }
{ (2\cosh \beta \Wh{h})^n }
\end{equation}
and $\Wh{\pi}(\Wh{h})$ is the distribution of the cavity bias, which is
related to $\pi(h)$ as
\begin{equation}
\Wh{\pi}(\Wh{h})=\int dh \pi(h)\left[\delta\left(
\Wh{h}-\frac{1}{\beta}
\tanh^{-1}(\tanh \beta J \tanh \beta h) 
\right)\right].
\end{equation}
For $c=3$ in the limit $\beta n \to y$, we have 
\begin{eqnarray}
&&g^{(2)}(y)
=\log e^{y} \left (1-\frac{1}{2}(1-e^{-2y})(1-p_*)^2 \right )^3,\\
&&g^{(1)}(y)=\log \left (1-\frac{3}{4}(1-e^{-2y})(1-p_*)^2(1+p_*)\right ), \\
&&g_{\mu}(y)=g_{0}-\log\left(1-\frac{1}{2}(1-e^{-2y})(1-p_*)^2 \right ),
\end{eqnarray}
where $g_{0}=\log \int dh P(h) (2\cosh \beta h)^n$ 
is the contribution from a
boundary spin and $P(h)$ 
is the boundary-field distribution of the BL determined satisfying the condition
$\pi(0)=P(0)/(\int P(h) (2\cosh \beta h)^n) = p_*$. 

Equation (\ref{eq:BLdecomposition}) represents a distinctive feature of 
cycle free graphs. In most systems, the contribution from the boundary 
becomes negligible as the system size $N$ tends to infinity. 
However, eq.\ (\ref{eq:BLdecomposition}) indicates that 
such a contribution does not vanish for a BL 
since $r_{\rm B}\to (c-2)/(c-1)$ remains of the order of unity
even if $N=1+c(c-2)^{-1} \left ((c-1)^{L^\prime}-1 \right )$ becomes 
infinite. 
Nevertheless, the complete separation of contributions between 
the inside and the boundary in this equation 
implies that it is physically plausible to use
$g_{\rm I}(y)$, instead of $g_{N}^{\rm BL}(y)$,
in handling problems concerning the bulk part of the objective 
graph. Actually, such a replacement has been adopted in several
studies on cycle free 
graphs \cite{Meza2,Meza4}. 
In general, $g_{\rm I}(y)$ agree with $g(y)$ of an RRG, 
which provides the basis of the correspondence between BLs and RRGs. 

In spin-glass problems on cycle free
graphs,
the replacement of $g_{N}^{\rm BL}(y)$ with 
$g_{\rm I}(y)$ is crucial. 
To see this, we here investigate the large deviation properties of $[Z^n]$.
We denote the boundary condition as 
$P_{\rm B}(\V{h})=\prod_{i \in {\rm boundary}} \pi_i(h_i)$. 
Equation (\ref{eq:RZs3}) implies that $\Xi(y)$ is expressed as 
$\Xi(y)=\int d\V{h}P_{\rm B}(\V{h})
\exp\left (-yE_g(\V{h}) \right )$, where $E_g(\V{h})$ is 
the ground state energy when $\V{h}$ is imposed on the boundary. 
For general systems, including a BL, this yields the identity 
\begin{equation}
y^2 (\partial /\partial y)\left (y^{-1} g_N(y) \right )
=N^{-1} D(\widetilde{P}_{\rm B} |P_{\rm B}) \ge 0,
\end{equation} 
where $\widetilde{P}_{\rm B}(\V{h})=P_{\rm B}(\V{h})
\exp\left (-yE_g(\V{h}) \right )/\Xi(y)$ and 
$D(\widetilde{P}_{\rm B} |P_{\rm B})$ is the Kullback--Leibler
(KL) divergence between $\widetilde{P}_{\rm B}(\V{h})$ 
and ${P}_{\rm B}(\V{h})$. 
An implication of this relation from large deviation statistics is 
that the probability $P(f)$ that 
$E_g(\V{h})/N$ equals $f$
scales as $P(f) \simeq \exp \left (N \Sigma_{N}(f) \right)$
for large $N$, 
where $f$ and $\Sigma_{N}(f)$ are related by
$f=-(\partial /\partial y) g_N(y)$ and 
$
\Sigma_{N}(f)=-y^2 (\partial /\partial y)\left (y^{-1} g_N(y) \right )
$ 
parameterized by $y$. 
The non-negativity of the KL divergence indicates that 
the rate function 
$\Sigma_{N}(f)$ cannot be positive, 
which guarantees the normalization constraint $\int df P(f)=1$. 

\begin{figure}[htbp] 
\begin{center}
   \includegraphics[height=64mm,width=80mm]{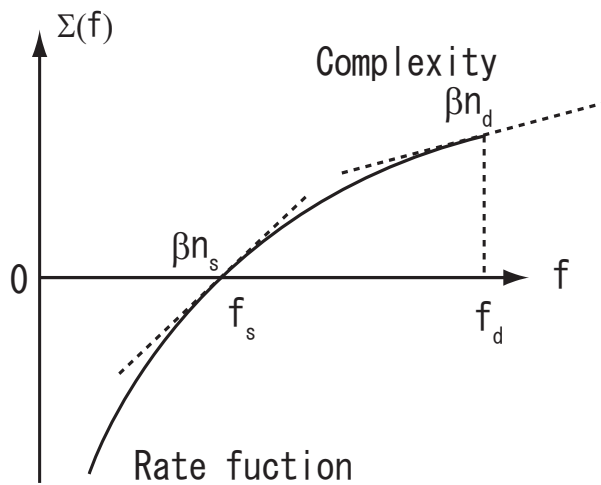}
 \caption{Schematic diagram of 
$\Sigma(f)$ assessed using $g_{\rm I}(y)$. 
The rate function is continued to the complexity 
at $f=f_{s}$ and the 1RSB occurs at this point. 
The complexity vanishes at $f=f_{d}$ where the monotonicity
of the free energy with respect to $y=\beta n$ breaks down. }
 \label{fig:Legendre}
\end{center}
\end{figure}

The constraint $\Sigma_{N}(f)\leq 0$ is always satisfied 
even when $N \to \infty$. However, this is not necessarily the case 
when we take the thermodynamic limit 
$\lim_{N \rightarrow \infty}g_{N}(y)=g(y)$ and 
then
calculate the rate function as
$\Sigma(y)=-y^2 (\partial /\partial y)\left (y^{-1} g(y) \right )$.
This function $\Sigma(f)$ can be positive, and    
it can be shown that the condition
$\Sigma(f_{s})=0$ signals the onset of $1$RSB \cite{Ogur1,Naka}. 
%This positive $\Sigma(f)$ can be formally interpreted as 
The positive part of $\Sigma(f)$ can be formally interpreted as 
the complexity or the configurational 
entropy of the metastable states 
%%%%%
for a {\em single} typical sample of couplings
%%%%
in the conventional 
1RSB framework \cite{Monn}, as shown in fig.\ \ref{fig:Legendre}. 
In the 1RSB framework, the critical condition 
$\Sigma(f_s)=0$, which is alternatively
expressed as $(\partial /\partial y) 
\left (y^{-1} g(y) \right )|_{y=y_{s}}=0$ in general, 
corresponds to the typical state realized in equilibrium. 

The condition $\Sigma(f_{s})=0$ has already been investigated
for RRGs and indicates that 1RSB transitions occur for some types
of RRGs \cite{Meza2,Mont}. However, 
it is considered that such a symmetry breaking cannot be detected by 
an investigation based on eqs.\
(\ref{eq:p0precise})--(\ref{eq:T0Zh2})
because 
the boundary contribution is inevitably
taken into account for a BL as well as for a CT. 
Actually, direct verification 
of $\Sigma(f) \leq 0$ is possible for the $c=3$ case; 
details are shown in \ref{sec:Sigma}. 
This indicates that the possible RZs provided by the 
current scheme are irrelevant to 1RSB. 

%%%%%%%%%%%%%%%%%%%%%%%%%%%%%%%%%%%%%%%%%%%%%%%%%%%%%%%%%%%%%%%%%%%%%%%%%%%%%%%%%%%%%%%%%%%%%%%%%%%%%%%%%%%%%%%%%%%%%%%%%%%%%%%%%%%%%%%%%%%%%%%%%%%%%%%%%%%%%%%%%%%%%%%%%%%%%%%%%%%%%%%%%%%%%%%%%%%%%%%%%%%%%%%%%%%%%%%%%%%%%%%%%%%%%%%%%%%%%%%%%%%%%%%%%%%%%%%%%%%%%%%%%%%%%%%%%%%%%%%%%%%%%%%%%%%%%%%%%%%%%%%%%%%%%%%%%%%%%%%%%%%%%%%%%%%%%%%%%%%%%%%%%%%%%%%%%%%%%%%%%%%%%%%%%%%%%%%%%%%%%%%%%%%%%%%%%%%%%%%%%%%%%%%%%%%%%%%%%%%%%%%%%%%%%%%%%%%%%%%%%%%%%%%%%%%%%%%%%%%%%%%%%%%%%%%%%%%%%%%%%%%%%%%%%%%%%%%%%%%%%%%%%%%%%%%%%%%%%%%%%%%%%%%%%%%%%%%%%%%%%%%%%%%%%%%%%%%%%%%%%%%%%%%%%%%%%%%%%%%%%%%%%%%%%%%%%%%%%%%%%%%%%%%%%%%%%%%%%%%%%%%%%%%%%%%%%%%%%%%%%%%%%%%%%%%%%%%%%%%%%%%%%%%%%%%%%%%%%%%%%%%%%%%%%%%%%%%%%%%%%%%%%%%%%%%%%%%%%%%%%%%%%%%%%%%%%%%%%%%%%%%%%%%%%%%%%%%%%%%%%%%%%%%%%%%%%%%%%%%%%%%%%%%%%%%%%%%%%%%%%%%%%%%%%%%%%%%%%%%%%%%%%%%%%%%%%%%%%%%%%%%%%%%%%%%%%%%%%%%%%%%%%%%%%%%%%%%%%%%%%%%%%%%%%%%%%%%%%%%%%%%%%%%%%%%%%%%%%%%%%%%%

\section{Results}\label{sec:Results}
\subsection{Plots of the replica zeros for tree systems}\label{sec:plots}
We here present the results only for CTs. RZs plots for ladder systems
are summarized in \ref{sec:BPladder}.
\begin{figure}[t]
\begin{tabular}{cc}
%\hspace{-5mm}
\begin{minipage}[t]{0.5\hsize}
\begin{center}
 \includegraphics[height=60mm,width=75mm]{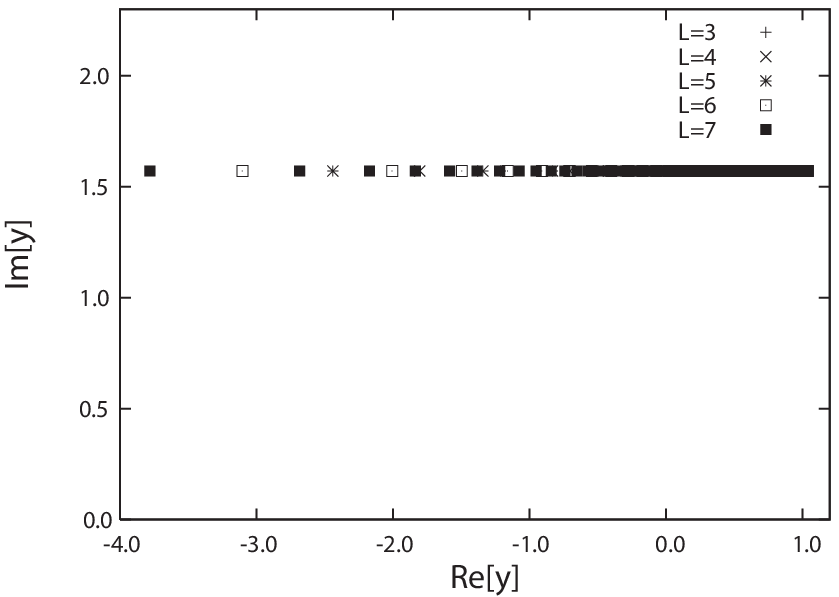}
 \caption{Plot of RZs for a
 CT with $c=3$. All the zeros 
lie on the line ${\rm Im} (y)=\pi/2$, as for a $2 \times L$ ladder. }
\label{fig:RBGzeros}
\end{center}
\end{minipage}
%\hspace{2mm}
 \begin{minipage}[t]{0.5\hsize}
\begin{center}
%\vspace{-5mm}
\includegraphics[height=60mm,width=75mm]{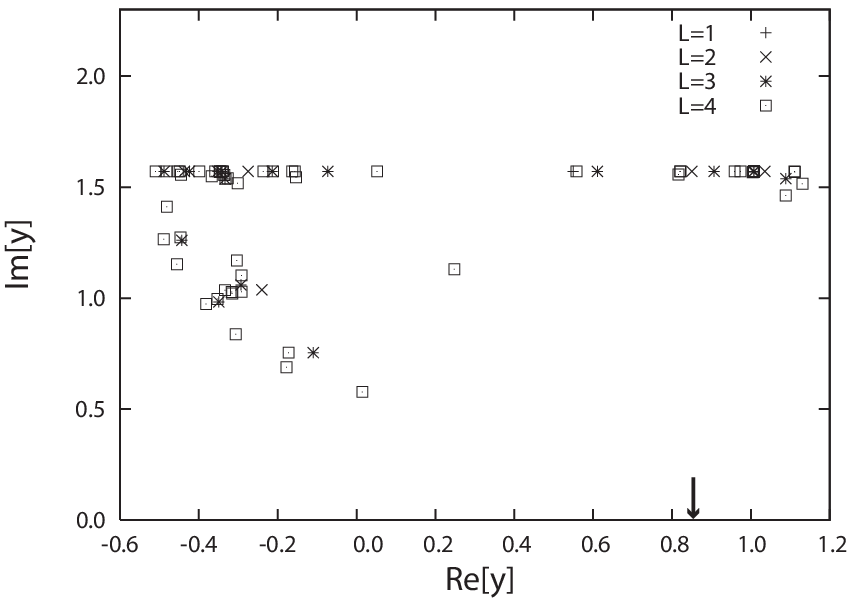}
\caption{RZs plot for a $3$-CT with $c=3$. 
A sequence of zeros approaches 
 the real axis as the number of generations $L$ increases. The arrow
 indicates the collision point expected from the study of the
 $L \to \infty$ limit in sec.\ \ref{sec:thermo}.}
\label{fig:3bRBGzeros}
\end{center}
\end{minipage}
\end{tabular}
\end{figure}
The plots for a CT and for a $3$-CT with $c=3$ are shown 
in figs. \ref{fig:RBGzeros} and \ref{fig:3bRBGzeros},
respectively. 
Note that 1RSB occurs in RRGs with the same parameters. 
The critical values are $y_s=0.41741$ and $\infty$
for the RRG counterparts of a CT and $3$-CT with $c=3$, respectively. 

Figure \ref{fig:RBGzeros} shows that 
RZs of the $c=3$ CT lie on a line ${\rm Im}(y)=\pi/2$.
Interestingly, this behavior is the same as 
the $2\times L$ ladder case, the plot of which is given 
in \ref{sec:BPladder}.
This result indicates that 
 there is no phase transition or breaking 
of analyticity of $g(n)$ with respect to real $y$.
This is in accordance with the argument on the boundary contribution 
mentioned in the previous section. 

\begin{figure}[t]
\begin{tabular}{cc}
\begin{minipage}[t]{0.5\hsize}
\begin{center}
\includegraphics[height=60mm,width=70mm]{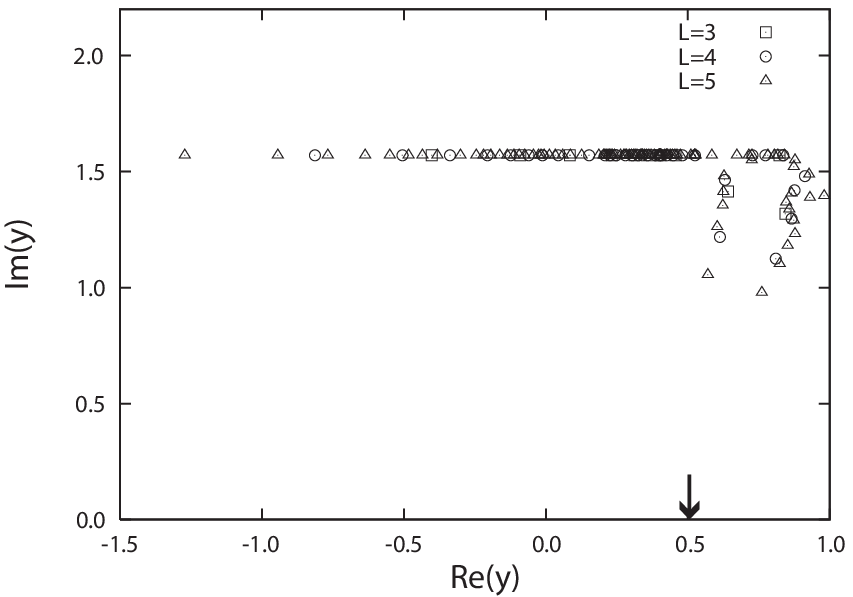}
 \caption{RZs of a CT with 
 $c=4$. We consider only an $L$-generation branch  in this case
 because of computational limits.
RZs approach the
 real axis as $L$ increases around $y_c \approx 0.5$. The arrow indicates the location of 
the singularity of the
 cavity-field distribution.}
 \label{fig:RBGzerosc4}
\end{center}
\end{minipage}
\begin{minipage}[t]{0.5\hsize}
\begin{center}
\includegraphics[height=60mm,width=70mm]{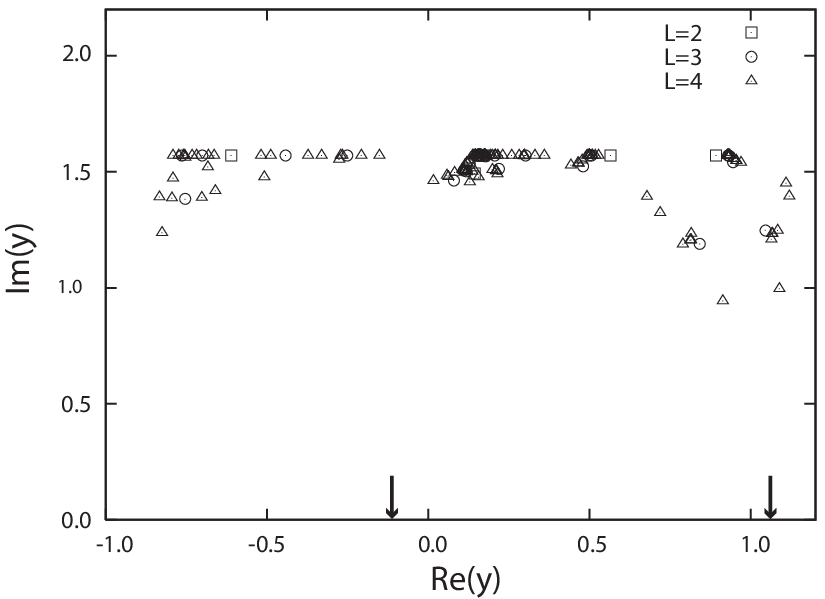}
   \caption{RZs of a $3$-CT with $c=4$. We consider only an $L$-generation branch. 
The zeros approach the real axis around $y_{c}\approx 1.1$. 
There are two singular points of the cavity field distribution in this
 case, both of which are indicated by arrows.}
   \label{fig:b3c4RBGzeros}
\end{center}
 \end{minipage}
\end{tabular}
\end{figure}

On the other hand, for the $3$-CT case in fig. 
\ref{fig:3bRBGzeros}, 
a sequence of RZs approaches 
a point $y_c$ on the real axis from the line 
${\rm Im}(y)=\pi/2$ as the number
of generations $L$ increases, although the value of $y_{c}$ is 
far from $y_{s}=\infty$. 
A similar tendency is also observed
for a CT and $3$-CT with $c=4$, plots of which are presented in
figs. \ref{fig:RBGzerosc4} and \ref{fig:b3c4RBGzeros}, respectively.
The 1RSB critical values are $y_{s}=0.38926$ for the CT
and $y_{s}=1.41152$ for the 3-CT. Again, these values are far from the values
of $y_{c}$, which can be observed 
in figs. \ref{fig:RBGzerosc4} and \ref{fig:b3c4RBGzeros}.

These results 
indicate that certain phase transitions occur for some CTs,
although they are irrelevant to 1RSB. 
It is difficult to identify the critical value $y_{c}$ 
from the plots because of the computational limits. 
Instead, 
in the following subsection we investigate the
$L \to \infty$ limit of these models. 
The arrows in figs. \ref{fig:3bRBGzeros}--\ref{fig:b3c4RBGzeros} 
represent the transition
points $y_{c}$ determined by this investigation.

\subsection{Phase transition on the boundary of a BL}\label{sec:thermo}
\begin{figure}[t]
%\hspace{-5mm}
\begin{tabular}{cc}
\begin{minipage}[t]{0.5\hsize}
\begin{center}
   \includegraphics[height=50mm,width=60mm]{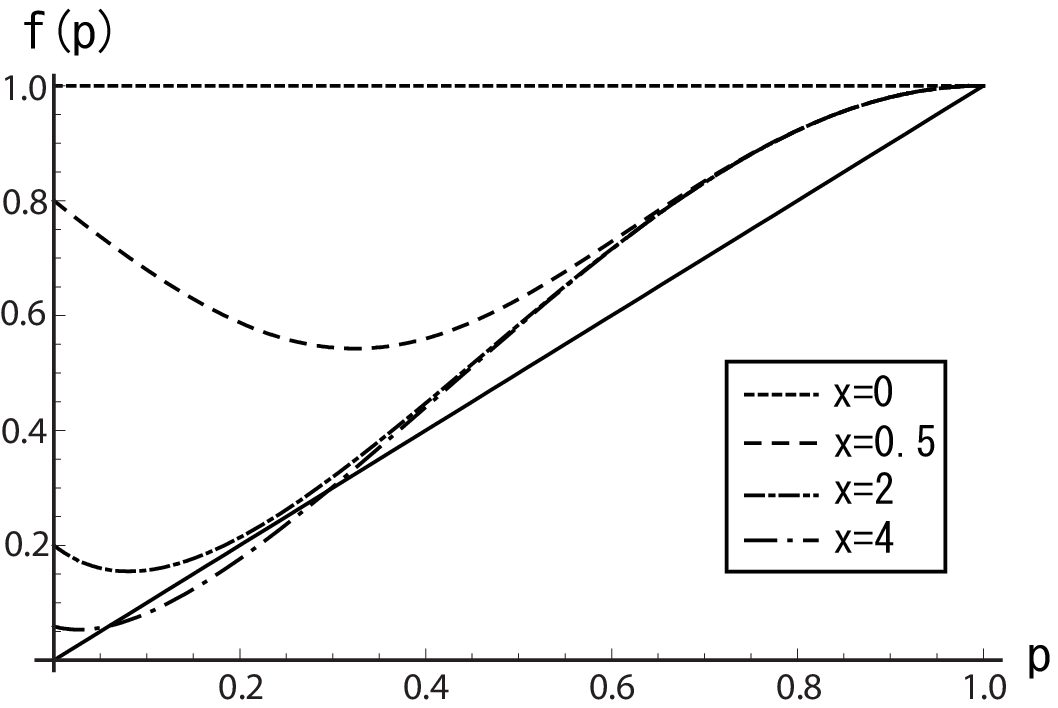}
 \caption{Return map of a $3$-CT with $c=3$. The
 convergent point of the recursion discontinuously changes depending on
 $x$. The solid line represents the function $f(p)=p$.}
 \label{fig:rmb3c3}
\end{center}
\end{minipage}
%\hspace{2mm}
\begin{minipage}[t]{0.5\hsize}
\begin{center}
   \includegraphics[height=50mm,width=60mm]{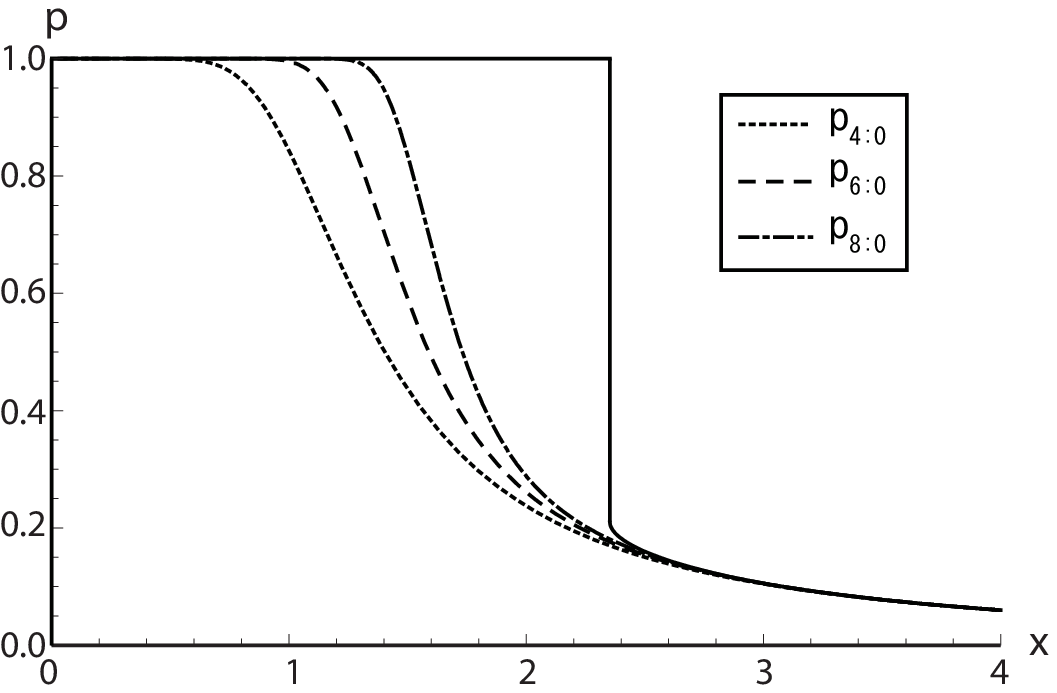}
 \caption{Asymptotic behavior of $p_{g;0}$ of a $3$-CT with
 $c=3$. A finite jump of $p$ occurs at $x\approx 2.35$. 
The solid line denotes the $\L \to \infty $ solution $p_{*}$.}
 \label{fig:apb3c3}
\end{center}
\end{minipage}
\end{tabular}
\end{figure}

In order to identify the value of $y_c$, 
we take the limit $L \to \infty$ by equating 
$p_{g+1;0}$ and $p_{g;0}$ in the iterative equation of $p_{g;0}$, which
yields the boundary condition $p_*$ of the BL.  For
a $c=3$ 3-CT, the iterative equation 
is given by
\begin{equation}
p_{g+1;0}=\frac{\left\{p_{g;0}^2+2p_{g;0}(1-p_{g;0})\right\}^2+\frac{1}{2}e^{-2y}(1-p_{g;0})^4}{1-\frac{1}{2}(1-p_{g;0})^4(1-e^{-2y})}.
\end{equation} 
A return map of the recursion of $p_{g;0}$ and the convergent solution 
$p_{*}$ are presented in 
figs. \ref{fig:rmb3c3} and \ref{fig:apb3c3}, respectively.
The return map shows that 
there are three fixed points 
for $x \gsim 2.35$, while 
$p=1$ is the only fixed point for $x \llsim 2.35$. 
This situation is in contrast to the $c=3$ CT case, in which 
the cavity-field distribution uniformly 
converges to an analytic function: 
\begin{equation}
p_{*}=\frac{2+x^2-\sqrt{x^4+8x^2}}{2(1-x^2)} \label{eq:pb2c3},
\end{equation}
which can be derived from eq.\ (\ref{eq:p0precise}). 
This implies that when eq.\ (\ref{eq:boundary_dist})
is put on the boundary of the CT, 
the boundary condition of the BL, which was obtained by 
an infinite number of recursions $L-L^\prime \to \infty$,
exhibits a discontinuous transition from $p_* < 1$ 
to $p_*=1$ at $x \approx 2.35 \Leftrightarrow y_c \approx 0.85$
as $y$ is reduced from the above.  
Actually, in fig.\ \ref{fig:3bRBGzeros}, 
RZs of the $c=3$ 3-CT seem to approach $y_c \approx 0.85$, 
marked by an arrow. 
This indicates that RZs obtained by our framework are 
relevant to the phase transition of the boundary of a BL, 
which is not related to 1RSB.    

\begin{figure}[t]
%\hspace{-5mm}
\begin{tabular}{cc}
\begin{minipage}[t]{0.5\hsize}
\begin{center}
   \includegraphics[height=50mm,width=60mm]{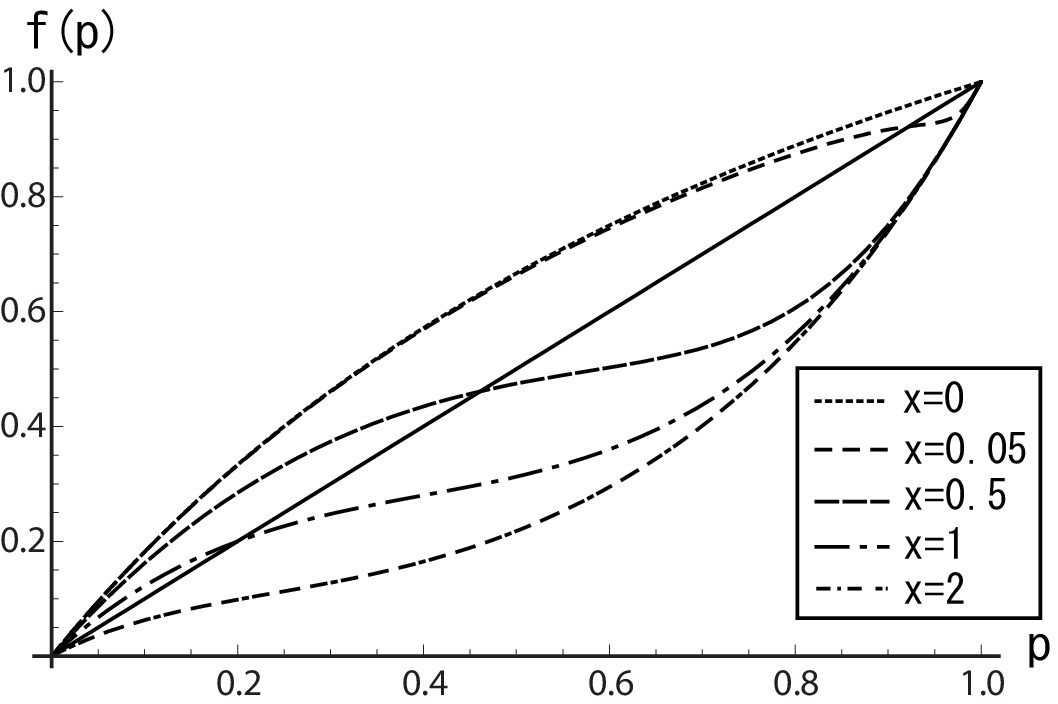}
 \caption{Return map of a CT with $c=4$. The
 stable fixed point is unique but shows a singularity at 
$x=e^y =\sqrt{3}$.}
 \label{fig:rmb2c4}
\end{center}
\end{minipage}
%\hspace{2mm}
\begin{minipage}[t]{0.5\hsize}
\begin{center}
   \includegraphics[height=50mm,width=60mm]{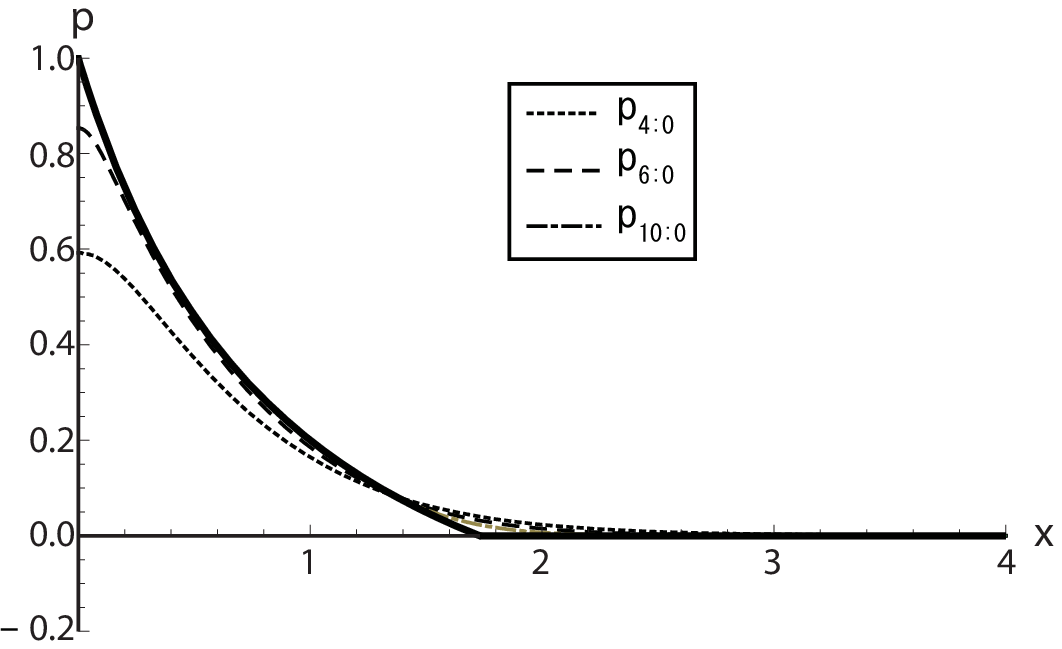}
 \caption{Asymptotic behavior of $p_{g;0}$ of a CT with $c=4$. In the thermodynamic limit, $p_{g;0}$ is continuous
 but the derivative becomes discontinuous at $x =\sqrt{3}$.}
 \label{fig:apb2c4}
\end{center}
\end{minipage}
\end{tabular}
\end{figure}

The same analysis for a $c=4$ CT shows that bifurcation 
of another type can occur for even $c$. 
For this model, 
the recursive equation of $p_{g;0}$ 
has a trivial solution $p_*=0$ for $\forall{x}$, 
which is always the case when $c-1$ is odd. 
The return map and plots of $p_*$ 
are shown in figs. \ref{fig:rmb2c4} and \ref{fig:apb2c4}, respectively. 

These figures indicate that  
there exists a continuous transition from $p_*=0$ to
$p_* >0$ at a certain value of $x$, 
which can be assessed as $x_{c}=\sqrt{3} \Rightarrow y_{c}\approx 0.5$.   
This is consistent with  
a certain sequence of RZs approaching 
the real axis around $y_c \approx 0.5$ in fig.\ \ref{fig:RBGzerosc4},
 which supports 
the analytical assessment of the critical points. 

In general, the discontinuous transition appears for 
cases of $k \ge 3$ spin interactions and 
the continuous transition occurs when $c$ is even. 
Actually, for a $c=4$ 3-CT, 
both discontinuous and continuous transitions
occur at 
$x \approx 0.86 \Rightarrow y \approx -0.15$ 
and $x=3 \Rightarrow y \approx 1.1$, respectively.
Figure \ref{fig:b3c4RBGzeros} shows 
a sequence of RZs approaching $y\approx 1.1$, 
%though it appears that
%no zeros approach the other critical point $y\approx -0.15$. 
while it is difficult to clearly identify a sequence converging to 
the other critical point $y\approx -0.15$. 
We consider that this is because the system size 
is not large enough, since a portion of the RZs in the left 
shows a tendency to approach the real axis,
though further increase of the system size 
is practically unfeasible due to the limitations of current 
computational resources. 

In conclusion, the analysis shown in this section indicates that 
RZs of CTs are related to the phase transitions on the boundary 
of a BL. Regardless of the type of transition, 
a sequence of RZs approaches a critical point on the real 
axis when the BL provided from a CT 
in the limit $L \to \infty$ exhibits a phase transition 
on the boundary. 

%%%%%%%%%%%%%%%%%%%%%%%%%%%%%%%%%%%%%%%%%%%%%%%%%%%%%%%%%%%%%%%%%%%%%%%%%%%%%%%%%%%%%%%%%%%%%%%%%%%%%%%%%%%%%%%%%%%%%%%%%%%%%%%%%%%%%%%%%%%%%%%%%%%%%%%%%%%%%%%%%%%%%%%%%%%%%%%%%%%%%%%%%%%%%%%%%%%%%%%%%%%%%%%%%%%%%%%%%%%%%%%%%%%%%%%%%%%%%%%%%%%%%%%%%%%%%%%%%%%%%%%%%%%%%%%%%%%%%%%%%%%%%%%%%%%%%%%%%%%%%%%%%%%%%%%%%%%%%%%%%%%%%%%%%%%%%%%%%%%%%%%%%%%%%%%%%%%%%%%%%%%%%%%%%%%%%%%%%%%%%%%%%%%%%%%%%%%%%%%%%%%%%%%%%%%%%%%%%%%%%%%%%%%%%%%%%%%%%%%%%%%%%%%%%%%%%%%%%%%%%%%%%%%%%%%%%%%%%%%%%%%%%%%%%%%%%%%%%%%%%%%%%%%%%%%%%%%%%%%%%%%%%%%%%%%%%%%%%%%%%%%%%%%%%%%%%%%%%%%%%%%%%%%%%%%%%%%%%%%%%%%%%%%%%%%%%%%%%%%%%%%%%%%%%%%%%%%%

\section{Discussion}
\subsection{Possible link to AT instability}\label{sec:DiscussAT}
The AT
condition, which is critical 
for FRSB, has not yet been characterized
for sparsely connected spin models. 
In fact, previous research has found that 
critical values of the continuous transitions
from $p_*=0$ to $p_* > 0$ 
are candidates for those of the AT condition for systems 
of even $c$ \cite{Mont}. 
This motivates us to further 
explore a possible link between RZs and the AT instability. 

Divergence of the spin-glass susceptibility of the root site $0$ 
is often adopted as the critical condition of 
the AT instability for BLs \cite{Thou,Rivo,Krza,Mart}. 
Generalizing the condition to the case of finite $n$, 
we obtain 
\begin{equation}
\chi_{SG} =\sum_{i}\left [
\left(\Part{\Ave{S_{0}}}{h_{i}}{} \right)^2 \right ]_n.
\label{eq:chiSG}
\end{equation} 
where $\left [ (\cdots )\right ]_n$ means 
an average with respect to a modified distribution of coupling and 
boundary field 
\begin{equation}
P_n(  \{J_{ij}\},\{h_{i}\}  )=
\frac{
P(  \{J_{ij}\},\{h_{i}\}   )Z^n(\{J_{ij}\},\{h_{i}\} )
}
{
\sum_{\{J_{ij}\}}
P(\{J_{ij}\},\{h_{i}\})Z^n(\{J_{ij}\},\{h_{i}\} )
}. 
\end{equation}
This definition is reasonable because eq.\ (\ref{eq:chiSG}) correctly 
reproduces the AT condition of fully connected systems
for finite $n$ in the limit of
infinite connectivity $c \to \infty$ \cite{Naka,Gard}.

In a cycle-free graph, an arbitrary pair of nodes is connected 
by a single path. Let us assign node indices from the origin 
of the graph $0$ to a node of distance $G$ along the path 
as $g=1,2,\ldots,G$. For a fixed set of couplings and boundary 
fields, the chain rule of the derivative operation 
indicates that
\begin{eqnarray}
\Part{\Ave{S_{0}}}{h_{G}}{}&=&\Part{\Ave{S_{0}}}{h_{0}}{}
\Part{h_0}{\Wh{h}_0}{}\Part{\Wh{h}_0}{{h}_1}{}
\cdots \Part{h_G}{\Wh{h}_{G}}{}
=\Part{\Ave{S_{0}}}{h_{0}}{}
\Part{h_0}{\Wh{h}_0}{} \prod_{g=1}^G
\Part{\Wh{h}_{g-1}}{h_{g}}{}
\Part{{h}_g}{\Wh{h}_g}{} \cr
&=& \Part{\Ave{S_{0}}}{\Wh{h}_0}{}
\prod_{g=1}^G
\Part{\Wh{h}_{g-1}}{\Wh{h}_g}{},
\label{chainrule}
\end{eqnarray}
as $h_g$ depends linearly on $\Wh{h}_g$ as 
$h_g=\Wh{h}_g+r_g$, where $r_g$ represents a sum of 
the cavity biases from other branches that flow into
node $g$. For a BL of $(k,c)=(2,3)$, the BP update yields 
an evolution equation of the cavity bias
\begin{eqnarray}
&&\Wh{h}_{g-1}=\frac{1}{\beta} 
\tanh^{-1}\left (\tanh(\beta J_g) \tanh(\beta (\Wh{h}_g+r_g))\right )\cr 
&&\to  \left \{
\begin{array}{ll}
{\rm sgn}\left (J_g (\Wh{h}_g+r_g) \right ) &(\,\, |\Wh{h}_g+r_g| \ge 1 \,\,) \cr
J_g (\Wh{h}_g+r_g)  &(\,\, \mbox{otherwise} \,\,) 
\end{array} \right .
(\,\,\beta\to \infty\,\,), 
\label{BPzerotemp}
\end{eqnarray}
where $J_g$ denotes the coupling between nodes $g-1$ and $g$, 
and similarly for other cases. 

To assess eq.\ (\ref{eq:chiSG}), we take an average of 
the square of eq.\ (\ref{BPzerotemp}) with respect to the 
modified distribution $P_n(\{J_{ij}\},\{h_{i}\})$. 
Here, $r_g$ can be regarded as a sample of a
stationary distribution determined by the convergent 
solution of eq.\ (\ref{eq:p0precise}) for the BL. 
As $r_g$ is limited to being an integer and $|J_g|=1$, 
eq.\ (\ref{BPzerotemp}) gives 
\begin{eqnarray}
\left |
\frac{\partial \Wh{h}_{g-1}}{
\partial \Wh{h}_g}
\right |=
\left \{
\begin{array}{ll}
0 & (\,\, |\Wh{h}_{g}+r_g|> 1 \,\,)\\
0 \mbox{ or $1$} & (\,\, |\Wh{h}_{g}+r_g|= 1\,\,) \\
1 & (\,\, {\rm otherwise}\,\,)
\end{array}
\right.,  \label{eq:transition}
\end{eqnarray}
where the value of $0$ or $1$ for the case of $|\Wh{h}_{g}+r_g|= 1$
is determined depending on the value of $\Wh{h}_{g}$.
%%%%%%%%%%%%%%%%%%%%%%%%%%
When $\Wh{h}_{g}=0$ (and $|r_g|=1$), the values $0$
and $1$ are chosen with equal probability $1/2$ since
the sign of the infinitesimal fluctuation of $\Wh{h}_{g}$, 
$\delta \Wh{h}_{g}$, 
is determined in an unbiased manner due to the mirror symmetry 
of the distribution of couplings. 
On the other hand, under the condition of $\prod_{k=g+1}^{G} \left |
\partial \Wh{h}_{k-1}/\partial \Wh{h}_{k} \right | \ne 0 $, 
the case of $|\Wh{h}_g|=1$ (and $r_g=0$) always yields $\left |
\partial \Wh{h}_{g-1}/\partial \Wh{h}_g \right |=1$. 
This is because $\Wh{h}_g \delta \Wh{h}_g <0$ is 
guaranteed for $|\Wh{h}_g|=1$ under this condition. 
%%%%%%%%%%%%%%%%%%%%%%%%%%

Equation (\ref{eq:transition}) indicates that 
the assessment of eq.\ (\ref{chainrule}) 
is analogous to an analysis of a random-walk 
which is bounded by absorbing walls. 
We denote by $P_{(G \to 0)}$ 
the probability that 
$\left |\partial \Wh{h}_{g-1}/\partial \Wh{h}_g\right |$ 
never vanishes during the walk from $G$ to $0$ and 
the value of 
$\prod_{g=1}^G
|\partial \Wh{h}_{g-1}/\partial \Wh{h}_g|$ is kept to unity. 
This indicates that 
\begin{eqnarray}
\left [\left (\Part{\Ave{S_{0}}}{h_{G}}{} \right )^2 \right ]_n
\propto 
P_{(G \to 0)}
\label{survingprob}
\end{eqnarray}
holds. Summing all contributions up to the boundary of the BL
yields the expression
\begin{equation}
\chi_{SG}  \propto 
\sum_{G=0}^{L^\prime} (k-1)^{G} (c-1)^{G} P_{(G \rightarrow 0)}. 
\label{expression_of_chiSG}
\end{equation}
The critical condition for convergence of eq.\ (\ref{expression_of_chiSG})
in the limit $L^\prime \to \infty$ is
\begin{eqnarray}
\log \left ((k-1)(c-1) \right )+\lim_{G \to \infty} \frac{1}{G} 
\log P_{(G \rightarrow 0)} =0.
\label{AT}
\end{eqnarray}
This serves as the ``AT'' condition in the current framework. 

For a BL, eq.\ (\ref{AT}) can be assessed by analyzing the 
random walk problem of eq.\ (\ref{eq:transition}), as
shown in \ref{sec:AT}. 
We evaluated the critical $y_{AT}$ values of eq.\ (\ref{AT}) 
for several pairs of $(k,c)$, 
shown in Table \ref{tab:yvalue} along with other critical values. 
\begin{table}[hbt]
    \begin{center}
\begin{tabular}{c|c|c|c}
\noalign{\hrule height 0.8pt}
$(k,c)$ & $y_{AT}$ & $y_{c}$ & $y_{s}$  \\
\hline
$(2,3)$ & 0.54397 & none & 0.41741 \\
\hline
$(2,4)$ & 0.89588 & $\log \sqrt{3}\approx 0.54931$ & $0.38926$  \\
\hline
$(3,3)$ & 1.51641  & 0.85545 & $\infty$ \\
\hline
$(3,4)$ & 1.35403  & $-$0.15082, $\log 3\approx 1.09861$ & 1.41152 \\
\noalign{\hrule height 0.8pt}
\end{tabular}
\end{center}
\caption{Relevant values of $y$. Note that each kind of $y$ is
 calculated using different models. 
The 1RSB transition point $y_{s}$ is for
 RRGs and $y_{AT}$ is for RRGs or BLs. The
 singularity of the cavity-field distribution $y_{c}$ is common for all the models.} \label{tab:yvalue}
\end{table}
These results show that the values of $y_c$, 
which signal the phase transitions of the boundary condition 
of the BL, agree with neither $y_{AT}$ or $y_{s}$, 
implying irrelevance of RZs to the replica symmetry breaking. 

The irrelevance of RZs to the AT instability may be interpreted as follows. 
We can link the spin-glass susceptibility to 
$g_N(n)$ in general by considering the following extension: 
\begin{eqnarray}  
&&N \Wt{g}_{N}( \V{F};m,n) =\cr
&&\log \left[\left( \Tr{} \exp\left( -\beta\sum_{a=1}^{m}H({S^{a}})
+\sum_{l=1}^{N} F_{l}\sum_{a<b}S_{l}^{a}S_{l}^{b}  
\right)\right)\left(\Tr{} e^{-\beta H}\right)^{n-m} \right], 
\label{eq:phitilda}
\end{eqnarray}
by breaking the replica symmetry 
introducing replica symmetric interactions among
$m$ out of $n$ replica systems with coupling $\V{F}=(F_1,F_2,\ldots,F_N)$.  
Obviously, $g_N(n)=\Wt{g}_{N}( \V{F}=0;m,n)$ and
$g_N(n)=\Wt{g}_{N}( \V{F};m=1,n)$ hold. 
Analytically continuing eq.\ (\ref{eq:phitilda})
to $n,m \in \mR$ and 
expanding the obtained expression around $\V{F} = 0$ for
$m \simeq 1$ yields
\begin{equation}
N\Wt{g}_{N}( \V{F};m,n)\approx N g_{N}(n)+
\frac{m-1}{2}\V{F}^{T}\Wh{\chi}_{SG}\V{F}+({\rm  higher\,\, orders}),
\label{eq:phitildaexpansion}
\end{equation}
where $\Wh{\chi}_{SG}=\left (
\left [\left(\Ave{S_{l}S_{k}}-\Ave{S_{l}} \Ave{S_{k}}
\right)^2 \right ]_{n} \right )$ represents the spin-glass susceptibility matrix. 

Equation (\ref{eq:phitildaexpansion}) implies that 
the divergence of the spin-glass susceptibility is linked to 
analytical singularities of $\lim_{N \to \infty} \Wt{g}_{N}( \V{F};m,n)$ 
for $m \ne 1$. However, for $m=1$, which corresponds to $g_N(n)$
examined in the present paper, it is 
difficult to detect the singularity because 
the factor $m-1$ with $\V{F}^{T}\Wh{\chi}_{SG}\V{F}$ 
makes the divergence of the spin-glass susceptibility 
irrelevant to the analyticity breaking of 
$g(n)=\lim_{N \to \infty}g_N(n)$. 
A possible solution is to consider systems of $m \ne 1$ in the framework
of 1RSB. 
However, an examination along this direction is 
beyond the scope of the present paper.

\subsection{Physical implications of the obtained solutions}
We concluded that bifurcations of the fixed point solutions
of the BP update correspond to phase transitions
of the boundary condition of a BL and are not relevant to either 1RSB or FRSB. 
Before closing this section, we discuss the physical 
implications of the obtained solutions.

A naive consideration finds that 
the solution of $p_{\infty;0}=p=1$ corresponds to a paramagnetic phase 
implying that any cavity fields vanish and therefore
all spin configurations are equally generated. 
Note that this phase is of the ground states 
in the limit $\beta \rightarrow \infty$ and is different from the usual
temperature-induced phase.
 
For finite $p<1$, relevant fractions of the spins can take any direction
without energy cost because the cavity field on the site is $0$. 
This implies that the ground state energy is highly degenerate, 
which means that this solution describes a RS spin-glass phase. 
Actually, it is easy to confirm that the following equality holds:
\begin{eqnarray}
q_{\mu \nu}=\frac{ [\Tr{} S^{\mu}_{g} S^{\nu}_{g} e^{-\beta \sum_{\mu} H^{\mu }}] }{[Z^n]}=\Tr{} S^{\mu}_{g}S^{\nu}_{g}\rho_{g}(\V{S})=1-p_{g;0}.
\end{eqnarray}  
Hence, the singularity of the cavity-field distribution in the limit 
$g\rightarrow \infty$ can be regarded as the transition of the spin-glass
order-parameter. A finite jump of $\pi_{\infty}(h)$ for the $k=3$ case is
the first-order transition from the RS spin glass to paramagnetic phases, and such
a transition is also observed in the mean-field models.  
The transitions from $p=0$ to finite-$p$ 
for the $c=4$
case can be regarded as a saturation of $q$ to $q_{EA}=1$. 
We infer that these are the transitions from RS to RS phases. 
Notice that such a transition has not been observed for 
infinite-range models.
Our results indicate that this $q=1$ phase appears 
only when $c$ is even.
This means that such a phase is highly sensitive to
the geometry of the objective lattice. This may be 
a reason why such a transition has not been observed in other models.

\section{Summary}
In summary, we have investigated RZs for CTs and
ladders in the limit $T,n\rightarrow 0,\,\, 
\beta n \rightarrow y \sim O(1)$. Most of the
zeros exist near the line ${\rm Im}(y)=\pi/2$ in all cases investigated; 
in particular, for the $(k,c)=(2,3)$
CT and the width-$2$ ladder all the zeros lie on this
line. For the width-$2$ ladder we have proved that 
the free energy is analytic with respect to $y$ in this model. 
On the other hand, for some CTs, a relevant fraction
of the RZs spreads away from the line ${\rm Im}(y)=\pi/2$ and
approaches the real axis as the generation number $L$ grows. 
This implies that $g(n)$ has a singularity at a finite real $y$ in
the thermodynamic limit.  
A naive observation finds that the RZs collision points 
correspond to phase transitions of the boundary condition
of the BL. We have compared them with known 
critical conditions of 1RSB and FRSB
and concluded that these conditions are irrelevant to the behavior of RZs. 
This is consistent with  
the absence of RSB in CTs
reported in some earlier studies.

To fully understand and use the replica method, 
as well as mathematical verification,
an description of the physical significance of the method is required. 
We hope that our results presented in this paper lead to a deeper
understanding of the mysteries of the replica method.

%\section*{Acknowledgement}
\ack
This work was supported by a Grand-in-Aid for Scientific Research on
the Priority Area ``Deepening and Expansion of Statistical Mechanical
Informatics'' by  the Ministry of Education, Culture, Sports, Science
and Technology as well as by CREST, JST.
We thank K. Hukushima and T. Nakajima  
for useful comments and discussions. 
We also acknowledge useful discussions with 
F. Ricci-Tersenghi and M. M\'ezard. 
\appendix

\section{Results for ladder systems}\label{sec:BPladder}
We first explain the procedure 
to obtain the RZs of ladder systems.
For a $2\times L$ ladder system, the BP equation can be derived 
in a similar manner to the CT case. 
\begin{figure}[t]
\begin{center}
   \includegraphics[height=40mm,width=50mm]{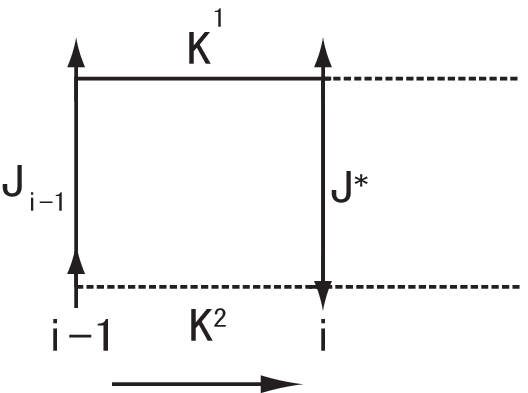}
   \caption{Unit cell of a $2\times L$ ladder.}
   \label{fig:w2RL}
\end{center}
\end{figure} 
We trace out the two spins 
of the previous generation to the $i$th generation, 
as in fig.\ \ref{fig:w2RL}. This 
yields an expression corresponding to
eq.\ (\ref{eq:basic}) as 
\begin{eqnarray}
&&\sum_{S_{i-1}^{1,2}}\exp\{\beta(J_{i-1}S_{i-1}^{1}S_{i-1}^{2}
+K^{1}S_{i-1}^{1}S_{i}^{1}+K^{2}S_{i-1}^{2}S_{i}^{2}
+J^{*} S_{i}^{1}S_{i}^{2}  ) \}\cr
&&=A\exp(\beta
(J^{*}+\Wh{J}_{i-1})S_{i}^{1}S_{i}^{2}  ).
\end{eqnarray}
From simple algebra we obtain
\begin{eqnarray}
&&\Wh{J}_{i-1}=\frac{1}{\beta}\tanh^{-1}(\tanh\beta K^{1} \tanh\beta K^{2} 
\tanh\beta J_{i-1}), \\
%%\,\, 
&&A=4\frac{\cosh\beta J_{i-1}\cosh\beta K^{1}\cosh\beta K^{2}}{\cosh\beta \Wh{J}_{i-1}}.
\end{eqnarray}
This shows that the effective bond $J_{i}$ between $S_{i}^{1}$ and
$S_{i}^{2}$ becomes
\begin{equation}
J_{i}=J^{*}+\Wh{J}_{i-1}.\label{eq:J}
\end{equation}
These relations indicate that the one-site marginal distribution
$\rho_{i}$ for trees is replaced by
the two-site marginal distribution $\rho_{i}(S_{i}^{1},S_{i}^{2})$ for a
width-$2$ ladder. From the symmetries of the original model, we can
specify the form of this distribution as
\begin{equation}
\rho(\V{S}_{i}^{1},\V{S}_{i}^{2})=\int dJ_{i} \pi(J_{i}) \frac{
e^{\beta J_{i} \sum_{\alpha} S_{i}^{1, \alpha }S_{i}^{2,\alpha}}  
}{(4\cosh \beta J_{i})^n}.
\end{equation}
This expression can be interpreted as showing that the effective bond 
fluctuates by quenched randomness. In a similar way to the tree case, 
the iterative equation for $\pi(J)$ is derived as
\begin{equation}
\pi_{i}(J_{I})\propto \int dJ_{i-1} \pi_{i-1}(J_{i-1})
\left[
\delta(J_{i}-J^{*}-\Wh{J}_{i-1})
\left(\frac{\cosh\beta J_{i}}{\cosh \beta \Wh{J}_{i-1}}\right)^n
\right],\label{eq:piJ}
\end{equation}
and that for the effective partition function is 
\begin{eqnarray}
&&\hspace{-10mm}
\Xi_{i}(n)=\Xi_{i-1}(n)\cr
&&\hspace{-10mm}\times\int dJ_{i-1}\pi_{i-1}(J_{i-1})\left[
(2\cosh\beta K^{1})^{n}(2\cosh\beta K^{2})^{n}
\left(\frac{\cosh\beta J_{i}}{\cosh \beta \Wh{J}_{i-1}}\right)^n
\right].
%&&=\Xi_{i-1}(n)\int dJ_{i-1}\pi_{i-1}(J_{i-1})\left[
%(2\cosh\beta J_{i-1})^{3n}
%\left(
%\frac{1+\tanh\beta J^{*}\tanh\beta \Wh{J}_{i-1}}{2}
%\right)^n
%\right].
\end{eqnarray}
In the limit $\beta \rightarrow \infty, \beta n \rightarrow y$, 
we can derive the following formulas from above equations:  
\begin{eqnarray}
&&p_{g+1;0}=\frac{1-p_{g;0}}{1-p_{g;0}-(1+p_{g;0})e^{2y}},\label{eq:T0piJ}\\
&&\Xi_{g+1}=\Xi_{g}e^{3y}\left(p_{g;0}+\frac{1}{2}(1-e^{-2y})(1-p_{g;0})\right)\label{eq:T0ZJ}.
\end{eqnarray}
Using these relations, we can symbolically calculate 
the $\Xi_{L}$ as a polynomial of $x=e^{y}$ and obtain the RZs by numerically 
solving $\Xi_{L}=0$ as the CT cases.

On the other hand, for larger-width ladders, 
the number of spins added by an iteration is greater
than $2$ and many-body interactions appear.
It makes the problem complicated and  
simple relations like (\ref{eq:J}) cannot
be obtained. 
Hence, when treating larger-width ladders,
we directly use the BP formula for a given sample $\{J_{ij}\}$ 
to obtain the ground state energy \cite{Kado}
and 
numerically 
assess the distribution of the ground state energies $P(E_{g})$
by enumerating all the configurations. 
Using the distribution $P(E_{g})$, the RZs equation is derived as 
\begin{equation}
\sum_{E_{g}} P(E_{g}) e^{-yE_{g}}=0.
\end{equation}
This equation is solved numerically in the same way as the other cases.
Note that computational times required in the counting process to obtain 
$P(E_{g})$ exponentially increases as $L$ grows, 
which makes it infeasible to obtain $\Xi_{L}$ for large $L$.

Next, we present the plots of RZs for ladder systems 
with brief discussions. 
\begin{figure}[t]
\begin{tabular}{cc}
%\hspace{-5mm}
\begin{minipage}[t]{0.5\hsize}
\begin{center}
   \includegraphics[height=60mm,width=75mm]{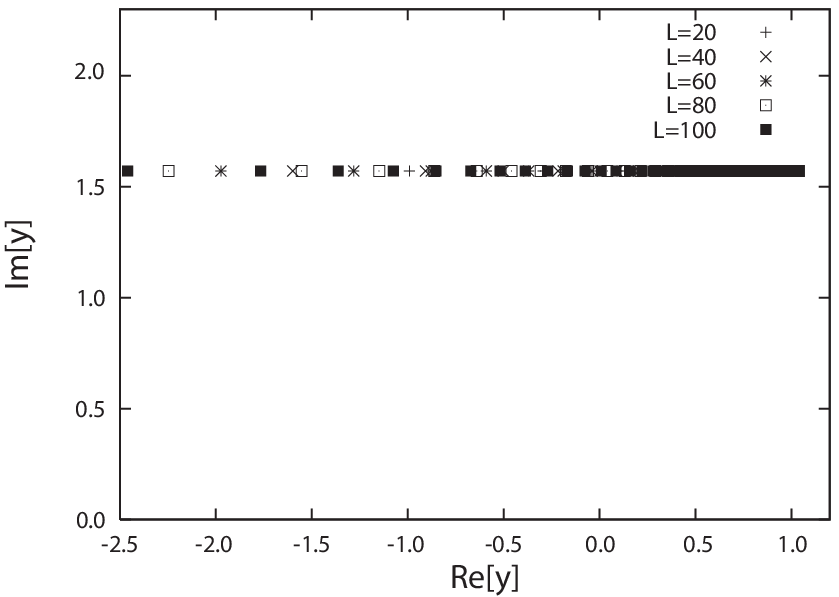}
 \caption{RZs for ladders with $2\times L$. All
 the zeros lie on ${\rm Im} (y)=\pi/2$ and never reach the 
real axis of $y=\beta n$. The inequality ${\rm Re}(y)\leq \log 2\sqrt{2}$
 holds, as shown in \ref{sec:proof}.}
 \label{fig:ladderzeros}
\end{center}
\end{minipage}
%\hspace{2mm}
 \begin{minipage}[t]{0.5\hsize}
\begin{center}
\includegraphics[height=60mm,width=75mm]{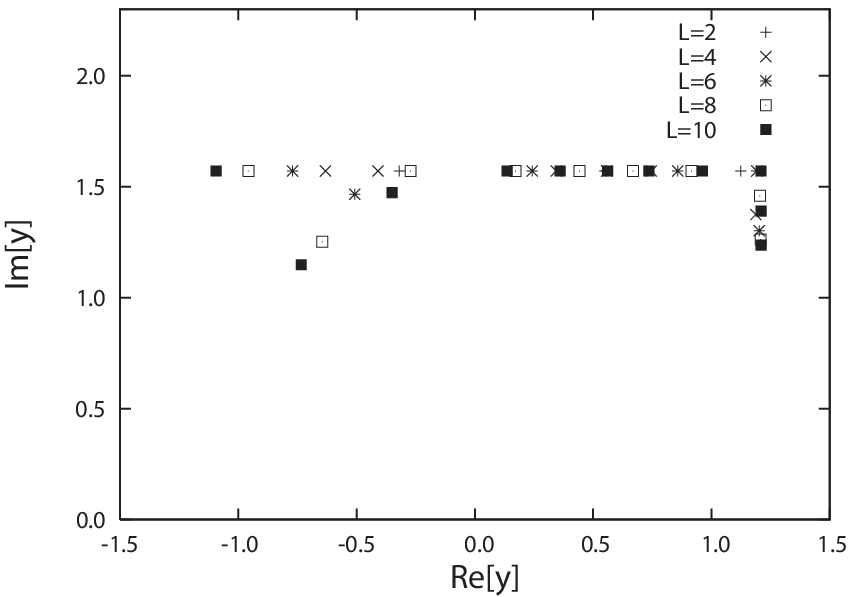}
\caption{Zeros of width-$4$ ladders. Some 
of the zeros approach  the real axis around ${\rm Re} (y)\approx 1.2$, but the
rate of approach rapidly decreases as $L$ grows.}
\label{fig:w4ladderzeros}
\end{center}
\end{minipage}
\end{tabular}
\end{figure}
Figure \ref{fig:ladderzeros} shows the plot for a $2 \times L$ ladder 
with the boundary condition $p_{0;0}=0$ and $\Xi_{0}=x$. 
Notice that all RZs lie on a line ${\rm Im}(y)=\pi/2$.
This fact can be mathematically proven, 
as detailed in \ref{sec:proof}. 
The physical significance of this behavior is
that the generating function $g_{N}(n)$ is analytic 
with respect to
real $y$ even for the $N \rightarrow \infty$ limit. 
We have also investigated a $3 \times L$
ladder and found qualitatively similar results as for the width-2 case.   
For a width-$4$ ladder,
the RZ plot is given in fig.\ \ref{fig:w4ladderzeros}. 
We can observe that some zeros approach the real axis around ${\rm Re}
(y)\approx 1.2$, 
but the rate of approach decreases rapidly 
as $L$ grows. This implies that the RZs do not reach the
real axis, which agrees with a naive
speculation that ladders are essentially one-dimensional systems and 
therefore do not involve any phase transitions as long as the 
width is kept finite.

\section{Location of replica zeros of a width-$2$ ladder}\label{sec:proof}
We prove that 
all RZs of a $2\times L$ ladder lie on the line 
${\rm Im}(y) =\pi/2$ for any $L$. 
We introduce the notation 
\begin{equation}
p_{l}(x)=p_{l;0}=\frac{n_{l}(x)}{d_{l}(x)},
\end{equation}
where $d_{l}$ and $n_{l}$ are polynomials of $x=e^y$ and 
$n_{l}(x)/d_{l}(x)$ is assumed to be irreducible.
The outline of the proof is as follows. First we present the general
solution of $p_{l}$ and show that
the denominator $d_{l}$ has $2F((l+1)/2)$ roots which are all
purely imaginary. 
The function $F(l)$ is the floor function, which is defined to return 
the maximum integer $i$ in the range $i \leq l$.     
Also, we show that 
the number of nontrivial solutions of $\Xi_{l}=0$ is 
equal to $2F((l+1)/2)$ and
$\Xi_{l}$ can be factorized as $C_{l}(x)d_{l}(x)$, where $C_{l}(x)$ is
a polynomial of $x$. From the correspondence of the numbers of the roots, 
we conclude that all the zeros of $\Xi_{l}$ are 
equivalent to the roots of $d_{l}(x)$ and $C_{l}(x)$ takes the form $ax^{b}$. 
  
The iteration (\ref{eq:T0piJ}) for $p_{l}$ has 
a solvable form and its general solution is given by
\begin{equation}
p_{l}=\frac{2(4^l-h(x)^l)}{4^{l}(2+x^2-x\sqrt{x^2+8})-(2+x^2+x\sqrt{x^2+8})h(x)^l},\label{eq:p_l}
\end{equation}
where
\begin{equation}
h(x)\equiv -4-x(x+\sqrt{x^2+8})=4\frac{x+\sqrt{x^2+8} }{x-\sqrt{x^2+8}}.\label{eq:h(x)}
\end{equation}
The roots of the numerator in eq.\ (\ref{eq:p_l}) can be easily calculated as 
\begin{equation}
x=\left\{
		    \begin{array}{cc}
		     \pm 2\sqrt{2}i  \hspace{2cm} (l=2m+1) \\
		 \hspace{-0.5cm}  0, \pm 2\sqrt{2}i \hspace{2cm} (l=2m)
		    \end{array}
			  \right.\label{eq:nroots},
\end{equation}
where $i$ denotes the imaginary 
unit and $m$ is a natural number. Then, we concentrate on finding 
the roots of the denominator in eq.\ (\ref{eq:p_l}) 
except for those of the numerator (\ref{eq:nroots}). 
From numerical observations in sec. 3, 
we found that any of the roots $x^{*}$,
which satisfy
$\Xi_{l}(x^{*})=0$, are purely imaginary and bounded 
by $|x^{*}|\leq 2\sqrt{2}$. Hence, we assume these conditions  
and perform the variable transformation
$z=-xi$. Equating the denominator of eq.\ (\ref{eq:p_l}) to $0$, we get 
\begin{equation}
\left(\frac{h(-ix)}{4}\right)^l=\left(\frac{\sqrt{8-z^2}+iz}{\sqrt{8-z^2}-iz} \right)^l
=\frac{2-z^2-i\sqrt{8-z^2}}{2-z^2+i\sqrt{8-z^2}}\label{eq:droots}.
\end{equation}
%%%%%%%%%%%%%%%%%
We now enumerate the
number of solutions under conditions that $z$ is real and bounded as $-2\sqrt{2} \leq z \leq 2 \sqrt{2}$. Under these conditions, we can transform eq. (\ref{eq:droots}) into a simple form by using the polar representation. The result is 
\begin{equation}
e^{i(2\theta_{1}-\pi)l}=e^{i2\theta_{2}},\label{eq:droots2}
\end{equation} 
where  
$\sqrt{8-z^2}+iz=r_{1}e^{i\theta_{1}}\,\,(-\pi<\theta_{1}\leq\pi)$
 and 
$r_{2}e^{i\theta_{2}}=2-z^2-i\sqrt{8-z^2}\,\,(-\pi<\theta_{2}\leq\pi)$.
%%%%%%%%%%%%%%%%%%
While $z$ varies from
$-2\sqrt{2}$ to $2\sqrt{2}$ continuously, the radius $r_{1}$ stays at
a constant $2 \sqrt{2}$ and the argument $\theta_{1}$ varies
from $-\pi/2$ to $\pi/2$ in the positive direction. In the same
situation, $\theta_{2}$ changes from $+\pi$ to $-\pi$ in the negative
direction. The radius $r_{2}$ is not constant, but is finite in this 
range.
The variables $\theta_{1}$ and $\theta_{2}$ are obviously 
continuous and monotonic functions of $z$.      
Therefore, the argument of the left-hand side of eq.\ (\ref{eq:droots2})
starts from $\theta=0$ and rotates with angle $2l\pi$ 
in the positive direction and
the counterpart of the right-hand side varies from
the same point $\theta=0$ to $-4\pi$. 
This means that there are $l+1$ values of $z$ where the
factor $(2\theta_{1}(z)-\pi)l$ becomes equal to $2\theta_{2}(z)$ 
except for
trivial solutions $z=\pm 2\sqrt{2}$. When $l$ is even, 
these solutions contain a trivial solution $z=0$, which can also be 
confirmed from eq.\ (\ref{eq:p_l}). 
Hence, the number of nontrivial roots
of $d_{l}$ becomes $l+1$ for odd $l$ and $l$ for even $l$, which is
equivalent to $2F((l+1)/2)$. 

As already noted,
 the number of nontrivial solutions of $\Xi_{l}=0$ is equal
to $2F((l+1)/2)$. This can be understood by considering 
that the number of terms of $[Z^n]$ is
 determined by the maximum number of defects $n_{d}$.
In the $2\times l$ ladder case, the value of $n_{d}$ is given by 
$F((l+1)/2)$ and the number of terms is $n_{d}+1$. The highest degree of
the relevant polynomials for RZs comes from the difference between the
highest and lowest ground-state energies and is  
given by $2n_{d}=2F((l+1)/2)$, which 
yields the number of nontrivial solutions of $\Xi_{l}=0$.

Finally, we prove that $\Xi_{l}$ takes the form $A_{l}x^{b_{l}} d_{l}(x)$ by
 induction. From eqs.\ (\ref{eq:T0piJ}) and (\ref{eq:T0ZJ}) with the
 initial conditions $p_{0;0}=0,\,\,\Xi_{0}=x$, 
we derive 
\begin{equation}   
p_{1}=\frac{1}{x^2+1},\,\, \Xi_{1}= \frac{1}{2}x^2 (x^2+1),
\end{equation}
which satisfies the desired form. Assuming that 
the condition $\Xi_{l}=Ax^{b_{l}} d_{l}(x)$ 
is true for $l=k$, we substitute this expression into eq.\ (\ref{eq:T0ZJ})
to get 
\begin{eqnarray}
&&\Xi_{k+1}=Ax^{b_{k}}d_{k}x^3\left\{
\frac{n_{k}}{d_{k}}+\frac{1}{2}\left(1+\frac{1}{x^2}
\right) \left(1-\frac{n_{k}}{d_{k}}\right)
\right\}\cr
&&=\frac{1}{2}Ax^{b_{k}+1}\left\{
(x^2-1)n_{k}+(1+x^2)d_{k}\right\}.\label{eq:Xi_{k+1}}
\end{eqnarray}
Equation (\ref{eq:T0piJ}) can be written as  
\begin{equation}
p_{k+1}=\frac{d_{k}-n_{k}}{(x^2-1)n_{k}+(1+x^2)d_{k}}=\frac{n_{k+1}}{d_{k+1}},
\end{equation}
which gives
\begin{equation}
(x^2-1)n_{k}+(1+x^2)d_{k}=c_{k+1}(x)d_{k+1}(x),
\end{equation}
where $c_{k+1}$ is a polynomial and satisfies
 $c_{k+1}=(d_{k}-n_{k})/n_{k+1}$. Substituting this relation,
 we can rewrite eq.\ (\ref{eq:Xi_{k+1}}) as
\begin{equation}
\Xi_{k+1}=\frac{1}{2}Ax^{b_{k}+1}c_{k+1}(x)d_{k+1}(x).
\end{equation}
As we have already shown, the number of nontrivial zeros of
$\Xi_{k+1}$ is equal to that of $d_{k+1}$. This means that $c_{k+1}$ cannot have
nontrivial roots and hence $c_{k+1}$ takes the form
$Ax^b$. This completes the proof by induction and demonstrates our proposition that
all RZs for a $2\times L$ ladder have a constant imaginary 
part $i\pi/2$. 

\section{Rate function for a CT with $c=3$}\label{sec:Sigma}
We here calculate the generating function
$g_{L}(y)$ for finite $L$. 
Consider an $L$-generation branch of a $c=3$ CT. 
An explicit form $g_{L}(y)$ is
easily derived from eq.\ (\ref{eq:T0Zh1}) as 
\begin{equation}
g_{L}(x)=\frac{ 2^L }{ 2^{L+1}-1 } g_{0}
+\frac{ 2^{L+1}}{2^{L+1}-1}(1-2^{-L})\log x
+\frac{1}{4-2^{-L+1}}\sum_{i=0}^{L-1} \frac{\log f_{i}}{2^{i}},
\end{equation}
where $x=e^{y}$ and
\begin{equation}
g_{0}=\log \Xi_{0},\,\, f_{i}=f_{i}(x,p_{i;0})=1-\frac{1}{2}(1-x^{-2})(1-p_{i;0})^2,
\end{equation}
using the same notations as in sec.\ \ref{sec:formulation}.  
The rate function with finite generations $L$ is given by
\begin{eqnarray}
\Sigma_{L}(x)=\frac{2^L}{2^{L+1}-1}\left(g_{0}-x\log x \frac{d g_{0}}{d x}\right) \cr
+\frac{1}{4-2^{-L+1}} \sum_{i=0}^{L-1}\frac{1}{2^{i}f_{i}} \left(
f_{i}\log f_{i}-C_{i}(x)x \log x  
\right), \label{eq:sigmaL}
\end{eqnarray}
where the factor $C_{i}(x)$ is given by
\begin{equation}
C_{i}(x)=
\Part{f_{i}}{p_{i;0}}{} \frac{d p_{i;0}}{d x}+\Part{f_{i}}{x}{}
=(1-x^{-2})(1-p_{i;0}) \frac{d p_{i;0}}{d x} -x^3(1-p_{i;0})^2
.
\end{equation}

Let us denote $\Sigma_{\infty}(x)=\lim_{L \to \infty} \Sigma_{L}(x)$. 
Because the inequality $\Sigma_{\infty}\leq 0$ always holds, the 1RSB
transition does not occur as long as the condition
$\Sigma_{\infty}(x)=\Sigma(x)$ is satisfied. 

In the range $y \geq 0 \Leftrightarrow  1 \leq x$, 
the factor $f_{i}$ 
is bounded as $1/2\leq f_{i} \leq 1$. This guarantees the uniform convergence of $g_{L}(x)$. 
The boundedness of $(d p_{i;0}/d x)$ can also be shown 
with some calculations. 
These conditions guarantee that t
$\Sigma_{L}(x)$ converges to a function $\Sigma_{\infty}$ uniformly. 
Hence, from elementary calculus, 
the equality $\Sigma(x)=\Sigma_{\infty}(x)$ holds, which implies the
absence of 1RSB. 
The same conclusion is more explicitly
derived for a BL 
because $f_{i}$ does not depend on $i$. 

\section{AT condition for the $(k,c)=(2,3)$ case}\label{sec:AT}
We here evaluate the AT condition for a BL %or RRG 
with $(k,c)=(2,3)$.
To evaluate $P_{(G\rightarrow 0)}$, we construct the transition
matrix of our random-walk problem.  
For a given $(\Wh{h}_{g},\Wh{h}_{g+1})$, 
the posterior distribution of $r_g$ is given
as
\begin{equation}
p(r_{g}|\Wh{h}_{g})=p(r_{g},\Wh{h}_{g})/p(\Wh{h}_{g})
\propto e^{ y(|r_g+\Wh{h}_{g}|-|r_{g}|-|\Wh{h}_{g}|)}  p(r_{g}),
\end{equation}
where $p(r_{g})$ is the prior distribution of $r_{g}$. This enables us to
derive the concrete expression of $p(r_{g}|\Wh{h}_{g})$, summarized
in Table \ref{tab:pb}.
\begin{table}[hbt]
    \begin{center}
\begin{tabular}{c|c|c|c}
\noalign{\hrule height 0.8pt}
$ r_{g}$ $\backslash$ $\Wh{h}_{g}$& 1  & 0 & $-1$  \\
\hline
1 & 
 $\displaystyle \frac{1-p_{b}}{(1+p_{b})+(1-p_{b})e^{-2y}}$
 & $\displaystyle \frac{1-p_{b}}{2}$&
 $\displaystyle \frac{(1-p_{b})e^{-2y}}{(1+p_{b})+(1-p_{b})e^{-2y}}$ \\
\hline
0 & $\displaystyle \frac{2p_{b}}{(1+p_{b})+(1-p_{b})e^{-2y}}$ &
 $\displaystyle p_{b}$
 &$\displaystyle \frac{2p_{b}}{(1+p_{b})+(1-p_{b})e^{-2y}}$ \\
\hline
$-1$ & 
 $\displaystyle \frac{(1-p_{b})e^{-2y}}{(1+p_{b})+(1-p_{b})e^{-2y}}$
& $\displaystyle \frac{1-p_{b}}{2}$&
 $\displaystyle \frac{1-p_{b}}{(1+p_{b})+(1-p_{b})e^{-2y}}$ \\
\noalign{\hrule height 0.8pt}
\end{tabular}
\end{center}
\caption{Values of $p(r_{g}|\Wh{h}_{g})$ for $(k,c)=(2,3)$. The symbol $p_{b}$ is
 the probability that the cavity bias takes the value $0$. } \label{tab:pb}
\end{table}
We can distinguish three states of the walker at the $g$-step as follows:
\begin{description}
\item[$\ket{1}$:] The walker has already vanished. 
\item[$\ket{2}$:] The walker survives and $|\Wh{h}_{g}|=1$. 
\item[$\ket{3}$:] The walker survives and $|\Wh{h}_{g}|=0$.
\end{description}
Hence, using the relation (\ref{eq:transition}), 
the transition matrix $T$ can be written as
\begin{equation}
T=
 \left(      
   \begin{array}{ccc}
    1 & p_{1,1} & \frac{1}{2}p_{1,0} \times 2 \\
    0 & p_{0,1} & \frac{1}{2}p_{1,0} \times 2 \\
    0 & p_{-1,1} & p_{0,0}
	 \end{array}
 \right),
\end{equation}  
where $p_{r_{g},\Wh{h}_{g}}$ represents $p(r_{g}|\Wh{h}_{g})$ and  
the condition $p_{r_{g},\Wh{h}_{g}}=p_{-r_{g},-\Wh{h}_{g}}$ applies. 
When $|\Wh{h}_{g}+r_{g}|=1 $ and  $|\Wh{h}_{g}|=0$, 
the states $\ket{1}$ and $\ket{2}$ occur 
with equal probability $1/2$, 
while $\ket{2}$ is always 
chosen when $|\Wh{h}_{g}+r_{g}|=1 $ and $|\Wh{h}_{g}|=1$ as exlained 
in sec. \ref{sec:DiscussAT}.
This matrix has three eigenvalues: 
$\lambda_{1}=1,\,\lambda_{2},$ and $\lambda_{3}$.  
The eigenvector of the 
largest eigenvalue $\lambda_{1}=1$ corresponds to the state $\ket{1}$ or
the vanishing state. 
Hence, the surviving probability $P_{(G\rightarrow 0)}$ is given by 
$1-\braket{1}{G}$, where $\ket{G}$ is the state of the walker at the $G$ step.
For large $G$, the relevant state is of the second-largest eigenvalue
$\lambda_{2}$, and we get 
\begin{equation}
P_{(G \rightarrow 0)}\approx \lambda_{2}^{G}.
\end{equation}
Using the stationary solution (\ref{eq:pb2c3}), we obtain
$P_{(G\rightarrow 0)}$ as a function of $x=e^{y}$. 
The AT condition becomes 
\begin{equation}
\chi_{SG}\propto \sum_{G}(k-1)^G (c-1)^G P_{(G\rightarrow 0)} \rightarrow \infty
\Leftrightarrow 
(k-1)(c-1)\lambda_{2}>1.
\end{equation}
This condition is easily examined numerically and we can verify that 
the AT instability occurs at $y_{AT}\approx 0.54397$ for $(k,c)=(2,3)$.

\end{document}